\documentclass[sort, compress, 5p]{elsarticle}
\journal{Simulation Modelling Practice and Theory}

\newcommand{\mytitle}{DISSECT-CF: a simulator to foster energy-aware scheduling in infrastructure clouds}
\newcommand{\SMALLESTIMEGRANULARITYM}{\tau}
\newcommand{\SMALLESTIMEGRANULARITY}{$\SMALLESTIMEGRANULARITYM$}
\newcommand{\SIMULATIONTIMEM}{\mathcal{T}}
\newcommand{\SIMULATIONTIME}{$\SIMULATIONTIMEM$}

\usepackage{caption}
\usepackage[caption=false,font=normalsize,labelfont=sf,textfont=sf]{subfig}
\usepackage{amsmath}
\usepackage{amsfonts}
\usepackage{MnSymbol}
\usepackage{multirow}
\usepackage{algorithm}
\usepackage{algorithmic}
\usepackage{mdwlist}

\algsetup{indent=2.5mm}

\begin{document}

\begin{frontmatter}

\title{\mytitle}
\author[sztaki,uibk,iit]{Gabor~Kecskemeti\corref{cor1}}
\ead{kecskemeti.gabor@sztaki.mta.hu, gabor@dps.uibk.ac.at, kecskemeti@iit.uni-miskolc.hu}

\cortext[cor1]{Corresponding author. Tel: +36 1 279 6065}
\address[sztaki]{Laboratory of Parallel and Distributed Systems of the Institute for Computer Science and Control of the Hungarian Academy of Sciences~(MTA SZTAKI), Kende u. 13-17, Budapest 1111, Hungary}
\address[uibk]{Distributed and Parallel Systems Group of the Institute of Computer Science at the University of Innsbruck, Technikerstra{\ss}e 21a, Innsbruck 6020, Austria}
\address[iit]{Institute of Informatics at the University of Miskolc, Miskolc-Egyetemvaros 3515, Hungary}

\begin{abstract}
Infrastructure as a service (IaaS) systems offer on demand virtual infrastructures so reliably and flexibly that users expect a high service level. Therefore, even with regards to internal IaaS behaviour, production clouds only adopt novel ideas that are proven not to hinder established service levels. To analyse their expected behaviour, new ideas are often evaluated with simulators in production IaaS system-like scenarios. For instance, new research could enable collaboration amongst several layers of schedulers or could consider new optimisation objectives such as energy consumption. Unfortunately, current cloud simulators are hard to employ and they often have performance issues when several layers of schedulers interact in them. To target these issues, a new IaaS simulation framework (called DISSECT-CF) was designed. The new simulator's foundation has the following goals: easy extensibility, support energy evaluation of IaaSs and to enable fast evaluation of many scheduling and IaaS internal behaviour related scenarios. In response to the requirements of such scenarios, the new simulator introduces concepts such as: a unified model for resource sharing and a new energy metering framework with hierarchical and indirect metering options. Then, the comparison of several simulated situations to real-life IaaS behaviour is used to validate the simulator's functionality. Finally, a performance comparison is presented between DISSECT-CF and some currently available simulators.
\end{abstract}

\begin{keyword}Cloud Computing \sep Infrastructure as a Service \sep Energy-Awareness \sep Resource management \sep Simulation\end{keyword}

\end{frontmatter}

\section{Introduction}
Infrastructure as a Service~(IaaS) systems \cite{BERKELEYCLOUD,BreakClouds} build on virtualisation technologies to allow automated infrastructure provisioning. Virtual machine (VM) based provisioning gives users two major benefits: they do not need to be experts in physical infrastructure maintenance, and they can easily follow their demand patterns and scale their virtual computing infrastructure (composed of several VMs) with tools built on top of IaaS systems. These two benefits led to the wide and rapid adoption of such infrastructure offerings.

Unfortunately, their rapid adoption has led to infrastructures that still have plenty of open research issues (e.g., energy aware VM placement, service level objective specifications). However, even IaaS systems operated by academia are used in production nowadays. As production level systems are used by a multitude of users on a daily basis, changing the internal behaviour of such systems might hinder their user experience (such as reliability and usability). Thus, research focused on improving the internals of IaaS systems (e.g., introducing a new experimental virtual machine placement algorithm) cannot be done on such production systems directly. Consequently, to analyse new and novel ideas for internal behaviour, researchers are either restricted to severely limited IaaS deployments (e.g., rarely utilising more than a few hosts) or should resort to theoretical modelling of expected internal behaviour. However, results based on such research is often questioned by the operators of production clouds because their applicability to large scale systems is often not proven. Some researchers use simulators to further evaluate their models~\cite{SimuComp-AhmedSabayasachi,SimuCompare-zhao2012modeling}. These simulators allow researchers the evaluation of new ideas in life-like scenarios and as a result such simulators could pave the way for the new research results allowing their wide-spread adoption.

Although a plethora of IaaS related simulators exist even today, these simulators have very different focuses. Some are designed completely from the user's point of view and hide the cloud's internals so users can make decisions on how and what services should be moved to the clouds~\cite{SimGrid-Hirofuchi2013, iCanCloud-Nunez2011}. Because of their user orientation, in these simulators it is frequently problematic to introduce changes in IaaS behaviour. Some others~\cite{DCWorms-kurowski2013dcworms, GreenCloud-5683561} emphasise the need for precision for such simulations. Despite their extensibility, these simulators not only scale very poorly (making it problematic to evaluate more elaborate IaaS scenarios where sometimes thousands of physical machines collaborate), but they also require complex setup procedures to be precise (e.g., one should model every possible application in the system to receive realistic results). Finally, there are simulators that introduce some assumptions in the system that reduce the precision of the simulations but reach unprecedented speeds~\cite{CloudSim-calheiros2011CloudSim, SimGrid-Hirofuchi2013, GroudSim-ostermann2011groudsim}. Unfortunately, despite having clear advantages, they are  too specific to allow investigations on internal IaaS changes (e.g., GroudSim only models external interfaces of clouds, SimGrid merely focuses on virtualisation, and CloudSim has conflicting extensions -- e.g., power modelling is not available while using networking).

In this article, a new versatile simulation framework is presented (called DIScrete event baSed Energy Consumption simulaTor for Clouds and Federations -- DISSECT-CF). Compared to the previously mentioned simulators, DISSECT-CF offers two major benefits: a unified resource sharing model, and a more complete IaaS stack simulation (including for example virtual machine image repositories, storage and in-data-centre networking). The benefits of the sharing model are threefold: $(i)$ it allows a single framework to model resource bottlenecks (e.g., CPU, network), $(ii)$ generic resource sharing performance optimisations immediately improve entire simulations, $(iii)$ it provides a unified view on resource usage counters (i.e., allows resource type independent, generic monitoring). Finally, DISSECT-CF also opens up possibilities for more fine-grained energy consumption modelling by allowing the user to derive energy consumption from multiple resource usage counters. As a result of these new advancements, the new simulator could foster research on schedulers that could either have better insight into internal IaaS behaviour or collaborate with internal schedulers of IaaS systems in order to achieve previously unprecedented flexibility, adaptability and elasticity in future cloud systems.

Unfortunately, DISSECT-CF's focus on supporting research on infrastructure cloud schedulers introduces several limitations to its applicability. First of all, for performance reasons the simulator represents networks with a simple flow model, which has already been shown by several studies (e.g., \cite{SimGrid-velho2013validity}) to be inaccurate for smaller-sized network transfers. Fortunately, smaller-sized network transfers have a negligible influence on scheduling decisions in most cloud related schedulers. Also, because scheduler focused research usually uses task or virtual machine instantiation/termination traces for behavioural studies, DISSECT-CF uses the black box philosophy for applications. Thus, the simulator will not provide accurate results on resource utilisation if a particular application's behaviour cannot be approximated with simple resource consumption metrics (e.g., when there is unstable CPU utilisation for extended periods of time). In fact, these limitations are present in most simulators~(except those that have packet level network simulations or employ more complex flow models -- see \cite{GreenCloud-5683561, SimGrid-velho2009accuracy}). Finally, as the new simulator is aimed at providing a framework for researchers to experiment with the internals of infrastructure clouds, the included scheduling mechanisms themselves are present only as examples for future work and they do not extend the scheduling related state-of-the art themselves.

The behaviour of DISSECT-CF was analysed by first validating it against the behaviour of a small-scale infrastructure cloud at the University of Innsbruck. According to the findings of this article, the system's simulated behaviour matches real-life experiments with negligible error (in terms of  application execution time, larger scale network transfers and energy consumption). For larger scale experiments, DISSECT-CF was validated with proven results from two other simulators that are close to the new simulator's functionality (namely CloudSim \cite{CloudSim-calheiros2011CloudSim} and GroudSim \cite{GroudSim-ostermann2011groudsim}). Then, performance of these two simulators was compared to the newly proposed one. Comparisons were executed with both real-world~(using the Grid Workloads Archive~\cite{Iosup08thegrid}) and synthetic traces. The use of real-world traces also revealed that DISSECT-CF based simulations allow 1.5-32$\times$ faster behavioural analysis of simple cloud schedulers or VM placement strategies. The performance differences were further investigated through synthetic traces and it is shown that DISSECT-CF scales significantly better in complex resource sharing situations with the help of its unified resource sharing model (one can observe an improvement of even over 2800$\times$ in execution time in some cases).

The rest of this article is organised as follows. Section \ref{sec-relworks} presents the related research results. Then, Section \ref{sec-design} reveals the architecture of the newly proposed simulator and discusses its internal behaviour and extensibility options. Section \ref{sec-eval} analyses the properties of DISSECT-CF by comparing its behaviour to real-life systems and by comparing its performance to other simulators. Finally, Section \ref{sec-conclusion} concludes the article with a summary and with the identification of future research directions.

\section{Research background} \label{sec-relworks}
This section first reviews the scheduling scenarios that a cloud simulator might support. Then an overview is presented on the most popular cloud simulation platforms. Finally, the section concludes with a problem statement for the new simulator.

\subsection{Schedulers related to IaaS systems and their requirements on simulators} \label{sec-schedulers}

There are seven common kinds of schedulers that could have an influence on the behaviour of a virtual infrastructure created on top of IaaS cloud systems. In the following, a short overview is given of these kinds of schedulers with special attention on their requirements from a simulated environment. The list is presented from the schedulers that have the strongest user-side orientation to the most hidden schedulers in infrastructure systems. 
\begin{description*}
\item[Task to VM assignment.] If a user has large enough resource demands, then its virtual infrastructure might include multiple virtual machines that could host a particular kind of task. In such cases, whenever a new task arrives, the user has to decide on which virtual machine it should actually run the task. The decision can be automated with a scheduler and a queuing  system (similar to local resource managers -- e.g.,~\cite{Sobie:2013:HSC:2465848.2465850, 6332105}). In order to support research on these kinds of schedulers, \emph{simulators should be able to provide past and present VM level performance metrics} (e.g., temporal performance degradation of the VM's computing capabilities).
\item[Virtual infrastructure scaling.] When a user's resource demands are more dynamic and sometimes unpredictable, then he/she would frequently face heavy under- or over-utilisation of his/her virtual infrastructure. To better meet the demands of the newly arriving tasks, the virtual infrastructure should be able to automatically scale. This scaling is often achieved with a special scheduler (e.g., \cite{rodrigez2009,Sotomayor:2008:CBE:1383422.1383434}) that decides when to instantiate/terminate a particular kind of VM. Research on such schedulers need \emph{simulators that are capable of providing accurate VM management metrics} (e.g., virtual machine instantiation time).
\item[Cross-cloud VM allocation.] Some users have access to multiple cloud infrastructures. For such users, a new scheduler is needed (e.g., \cite{Tordsson2012358, 5279590}) which can choose between various cloud providers and dispatch VMs to them. The selection procedure is expected to take into account the availability, reliability and similar metrics of the various providers and also it should consider issues like placing processing close to big data. For such schedulers, \emph{simulators are required to offer infrastructure provider level metrics and data locality information.}
\item[VM placement.] Inside IaaS systems, user requests are no longer represented as tasks but they are only seen as VMs. As IaaS systems are highly automated, decisions to place a particular VM on a physical machine must be also done by a scheduler. This kind of scheduler (e.g., \cite{tsakalozos2011,6009246}) could have two main tasks: $(i)$ for already existing virtual machines, a new VM to host mapping could be identified which would allow a VM arrangement that considers both the VMs actual load and the providers current needs, and $(ii)$ for newly requested virtual machines, the scheduler should determine the host where the VM could be run. As these schedulers have diverse tasks, \emph{simulators should have the capability to disseminate the load of currently running virtual machines and also the utilisation of physical machines}.
\item[Physical machine state schedule.] Energy conscious IaaS systems aim at reducing their energy consumption in several ways. A simple way to do so is to consolidate VM load to the most energy efficient machines and switch the rest to a more energy efficient state. The automated decisions on which machines should be serving VMs and which ones should be waiting in low power states (e.g., suspend, switch off) are done by physical machine schedulers (e.g., \cite{5958802,6253507,Wang20131661}). These schedulers should ensure that, because of their operations, virtual machine creation and quality of service do not degrade below certain levels. To support the development of physical machine (PM) schedulers, \emph{simulators must necessarily maintain the cost of PM power state changes} (e.g., cold/suspend to RAM boot-up procedures).
\item[VM resource share management.] Schedulers are also present in virtual machine monitors~(like Xen or kvm) in order to allocate physical resources to virtual machines on a time-sharing basis (e.g., \cite{Ongaro:2008:SIV:1346256.1346258,5289182}). Although, these schedulers are not the main focus of research in cloud computing, they could have a direct impact on the quality of service if the above-mentioned VM placement strategies under-provision some virtual machines. For this reason, \emph{simulators should be able to correctly handle and report under-provisioning scenarios on physical machines}.
\item[Virtual resource assignment to task.] The lowest levels of schedulers that may affect higher-level (e.g., task to VM assignment) decisions are the process schedulers (e.g.,~\cite{Wong:2008:TAF:1400097.1400102, 4510751}) of the operating system in the user's VMs. In some cases the user could have an influence on the OS scheduler, but in others users must use OSs and schedulers that are prepared and accredited by the IaaS providers. Since these schedulers are generic OS level schedulers, they are out of the scope of cloud computing research. But since higher-level schedulers might make decisions on how these process schedulers behave, simulators should give their users some information on their behaviour. For example, \emph{simulators should be capable of reporting if a particular VM is under-provisioned} and tasks have no chance to access resources scheduled for them by the OS level scheduler.
\end{description*}

\subsection{Cloud simulators}

CloudSim~\cite{CloudSim-buyya2009modeling} is amongst the most popular IaaS cloud simulators. It was initially based on GridSim~(a widely used grid simulator developed by the same research institute~--~\cite{GridSIM-buyya2002gridsim}) but, after some performance and reliability issues, it was completely rewritten so it uses only some concepts (e.g., Cloudlet -- Gridlet analogy) from its predecessor~\cite{CloudSim-calheiros2011CloudSim}. CloudSim introduced the simulation of virtualised data centres mostly focusing on computational intensive tasks and data interchanges between data centres. Later, they extended the simulation to better support internal network communications of a data centre with NetworkCloudSim~\cite{CloudSim-garg2011networkCloudSim}. There are also extensions that simulate the energy consumption behaviour of the physical machines in the data centre based on specpower benchmarks and on dynamic voltage and frequency scaling~\cite{CloudSim-beloglazov2012optimal, CloudSimExt-DVFS}. CloudSim also formed an ecosystem. Several third parties offer extensions on top of CloudSim. Some significantly change CloudSim behaviour (e.g., add performance improvements~\cite{CloudSimExt-li2012dartcsim}, add better support for inter-cloud operations~\cite{SIMIC-6531742, SIMIC-6550488}, implement new energy consumption models~\cite{CloudSimExt-shi2011energy}, or introduce SLA concepts into the simulation~\cite{CloudSimExt-SLA}), while others wrap CloudSim and provide additional functionality (like graphical user interfaces for teaching~\cite{CloudSimExt-jararweh2012teachcloud} or for analytics~\cite{CloudSimExt-wickremasinghe2010cloudanalyst,CloudSimExt-CloudReports}). Despite its wide use, CloudSim has several disadvantages: $(a)$ low performance for scheduling research where thousands of scheduling scenarios should be evaluated in a timely fashion, $(b)$ networking is simulated for tasks only (e.g., data centre operations that utilise the same network as user tasks -- like virtual machine image transfers -- are not simulated even though they could have significant effects on the user perceived network performance) and $(c)$ using multiple extensions at once is frequently not possible (e.g., advanced networking and energy consumption modelling are not usable together since one would need to have virtual machines that inherit behaviour both from \verb+PowerVm+ and from \verb+NetworkVm+ classes).

The SimGrid framework~\cite{SimGrid-casanova2001simgrid} is another widely used simulator for analysing the behaviour of distributed systems (e.g., grids, peer to peer systems). Its resource sharing simulation is one of the most detailed; for example, it contains one of the most accurate non-packet oriented network models~\cite{SimGrid-velho2009accuracy,SimGrid-velho2013validity}. This simulator's focus was not particularly on clouds for a long time but recently its developers introduced extensions for virtualisation (e.g., hypervisors or live migration~--~\cite{SimGrid-hirofuchi2013adding,SimGrid-Hirofuchi2013}). Because of its distributed systems and grid background the simulator is inefficient in IaaS cloud related situations. For example, this simulator stops at the virtual machine level, thus it would require significant effort to build a multi data centre/cloud simulation on top of it.

While CloudSim and SimGrid were heavily influenced by previous simulators for grids and distributed systems, for performance reasons they also make compromises on networking. To resolve such issues there are simulators like iCanCloud~\cite{iCanCloud-Nunez2011} and GreenCloud~\cite{GreenCloud-5683561} that are built on network simulators (e.g., OMNeT++ or NS2) to more accurately simulate network communications in cloud systems. Their efforts result in great accuracy if all IaaS components and applications are modelled correctly network-wise; otherwise, they just introduce serious performance penalties because of the packet level simulations without the expected accuracy. In addition to networking improvements, GreenCloud~\cite{GreenCloud-kliazovich2012greencloud} is also offering precise energy estimations for networking and computing components, while iCanCloud also offers a user oriented simulation which supports IaaS utilisation decision-making~\cite{iCanCloud-nunez2011design} on top of the regular IaaS related simulation functionalities~\cite{iCanCloud-nunez2012icancloud}.

Next, GroudSim -- a simulator developed at the University of Innsbruck~\cite{GroudSim-ostermann2011groudsim} -- was analysed. This simulator aims at performance while it encompasses cloud concepts in a grid simulator environment. The simulator is also integrated with the ASKALON workflow system~\cite{GroudSim-ostermann2011integration} so it can be used to evaluate behavioural changes of real-life scientific workflows in the case of changes in the computing environment. Although this simulator supports clouds, it does not provide implementation on the internals of IaaS systems (i.e., it provides a black box implementation), thus it is not suitable for research studies that involve the internals of cloud infrastructures. And although GroudSim supports both CPU and network resources, the networking implementation of GroudSim is one of the least developed ones amongst the reviewed simulators.

The above simulators focus more on the user related behaviour of data centres, but there is a class of cloud simulators which is more focused on supporting decisions related to data centre operations~(e.g.,~\cite{SPECI-sriram2009speci, DCSim-tighe2012dcsim, DCWorms-piatek2013dcworms}). So even though these simulators could be used for examining user related behaviour, their detailed implementation of data centre behaviour reduces their usability in this context. On the other hand, these simulators offer some unique features that might be useful for research on IaaS related schedulers. For example, SPECI~\cite{SPECI-sriram2009speci} is focused on offering a tool to analyse the scalability of IaaS toolkits that will support future data centres. Next, DCSim~\cite{DCSim-tighe2013towards} allows the analysis of new Virtual Machine management operations (like relocation). Finally, DCWorms~\cite{DCWorms-kurowski2013dcworms} provides a unique view on data centre energy efficiency, including the heating, ventilation and air conditioning (HVAC) system's airflow and high granularity resource (e.g., individual CPU, memory modules, network interfaces) energy modelling.

\paragraph{Problem statement} The analysis of the related work leads to the conclusion that existing simulators have many drawbacks for those who would like to investigate scheduling scenarios in IaaS systems. To fulfil the needs of such scheduling scenarios, the rest of the article reveals a new infrastructure simulator that provides better insights on infrastructure behaviour for schedulers while maintaining the scalability of past simulators.

\section{Design and internals of the simulator} \label{sec-design}

\begin{figure}[tb]
\centering
\includegraphics[width=\columnwidth]{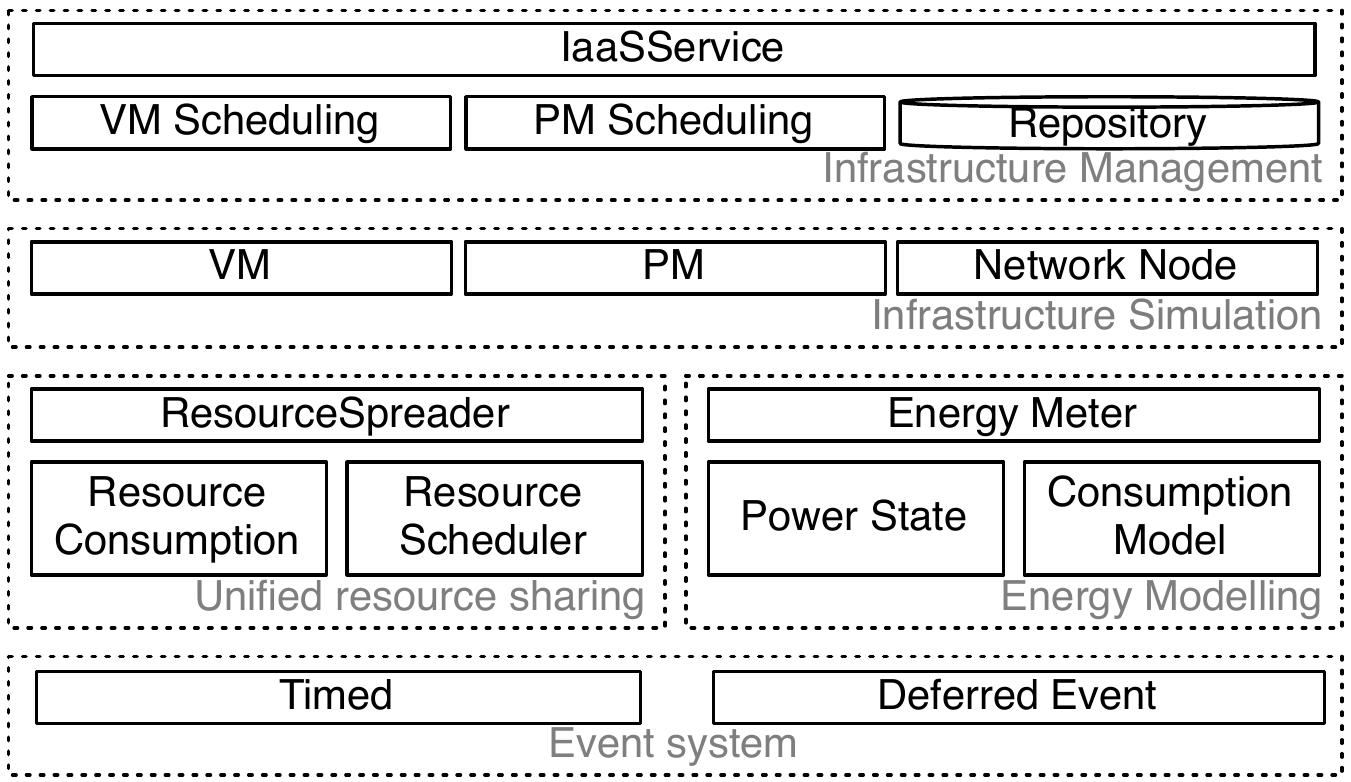}
\caption{Architectural view of DISSECT-CF \label{FIG-ARCH}}
\end{figure}

Figure~\ref{FIG-ARCH} presents the overall architecture of the newly proposed simulator\footnote{If not stated otherwise, the described algorithms, features and evaluation apply to DISSECT-CF version 0.9.5}. The figure groups the major components with dashed lines into subsystems. Each subsystem is implemented as independently from the others as possible.  As a result, simulation developers do not need to understand the complexity of the entire simulator if they intend to work on one of its subsystems. There are five major subsystems; they are listed in an order that follows their level of abstraction (from the most abstract to the more specific to IaaS systems):
\begin{description*}
\item[Event system.] These components provide the time reference for simulations.
\item[Unified resource sharing.]  \sloppy This subsystem acts as a lightweight and extensible foundation to low-level computing resource sharing (e.g., CPU, I/O).
\item[Energy modelling.] With these components DISSECT-CF enables simulator developers to monitor and analyse energy usage patterns of each individually simulated resource (e.g., network links, disks).
\item[Infrastructure simulation.] These components handle the behaviour of those IaaS system parts (e.g., virtual machines) that are the primary target of IaaS related schedulers.
\item[Infrastructure management.] This subsystem provides the user interface (the VM management API) and represents the high level functionalities (e.g., virtual machine schedulers) of infrastructure clouds.
\end{description*}
In the following sections, these subsystems are individually discussed.

\subsection{Event system}

The core of the DISSECT-CF simulator is a simple but high performance event generator~(reflected as \verb+Timed+ in Figure~\ref{FIG-ARCH}). It is used to maintain the time within the simulated system and allow third parties to be notified if a particular time instance has been reached.The simulator is not aware of the applied time granularity~(i.e., it is not known in the simulation if a single increase in the maintained time is equivalent to a single millisecond or a full hour). This enables flexibility in use, and allows simulation developers to have precision only when they assuredly need it; otherwise, they can benefit from faster simulations. In later sections of the article, the smallest time granularity for the current simulation is denoted with $\SMALLESTIMEGRANULARITYM\in\mathbb{R}$ and is expressed in seconds. Thus any given time instance in the simulator can be specified as: $t = \kappa\SMALLESTIMEGRANULARITYM$, where $\kappa\in\mathbb{N}$ and $t\in\SIMULATIONTIMEM$. Here, \SIMULATIONTIME refers to the set of all possible time instances throughout a simulation.

The simulator also assumes that notifications are recurring. Thus, subscribing to events means specifying the frequency with which one would like to be notified. The simulator contains a construct (called \verb+DeferredEvent+) for non-recurring events. Creating a subclass of either the \verb+Timed+ or the \verb+DeferredEvent+ classes allows simulation developers to receive custom time dependent notifications.

Finally, the \verb+Timed+ class is also the control point for the simulation time. Simulations have two distinct ways to influence simulated time:
\begin{description*}
\item[Instantaneous] controls directly influence the timer. First, one can \emph{fire} the events for the current time instance then advance the timer by one \SMALLESTIMEGRANULARITY. Second, it is possible to ask for a time jump that will progress the time with a given interval if within the interval there are no events expected.
\item[Continuous] controls let the simulation flow for a given time interval without any intervention~(e.g., one can simulate until all events from the queue are cleaned up). These controls also enable the progression of the timer while dropping irrelevant events that would occur in a given period of time.
\end{description*}

\subsection{A unified resource sharing model} \label{Sec-URSM}

Directly on top of basic time management lies the resource model of the simulator. The model is intended to capture low-level resource sharing behaviour (e.g., assigning tasks to virtual CPUs -- of VMs -- or virtual CPUs to physical ones, or balancing network bandwidth utilisation). DISSECT-CF applies a provider-consumer scheme to resources where resource consumptions are intermediaries between consumers and providers. In the case of simulated CPUs, consumptions represent instructions to be processed, thus CPU computing cycles of a physical machine are provided to virtual machines to consume.  In a network analogy, consumptions represent data to be transferred between two network hosts (where the sender acts as the provider and the receiver as the consumer).

\subsubsection{Foundations}

DISSECT-CF allows the definition of both providers and consumers with the help of the \verb+ResourceSpreader+ class~(see Figure~\ref{FIG-ARCH}). The set of all spreaders in a particular simulation will be referred as $\bar{S}$. The simulator uses the concept of resource consumptions as the intermediaries that represent the current processing demands of the actual consumers. Resource consumptions are denoted with a triplet: $c = <p_u,p_r,p_l>$, where $c$ represents the resource consumption, $p_u$ represents the processing that is currently under way, $p_r$ represents the remaining processing (i.e., processing that has not started yet) and $p_l$ represents the limit for this processing in a single \SMALLESTIMEGRANULARITY~(e.g., simulation developers can specify that a resource consumption is single-threaded so it can use the processing power of a single processor of a consumable CPU resource only). $\bar{C}$ represents all possible resource consumptions in a simulation: $c\in\bar{C}$. At a given time instance, the function $prov: \bar{C} \times \SIMULATIONTIMEM \to \bar{S}$ determines which provider offers the resources to be consumed. Similarly, $cons: \bar{C} \times \SIMULATIONTIMEM \to \bar{S}$ defines the consumer that utilises the offered resources. These functions are time dependent to allow the migration of resource consumptions  amongst spreaders.

At a given time, a particular resource consumption is processed in its provider by determining how much processing can be considered possible during a single \SMALLESTIMEGRANULARITY. The possible processing has an upper bound of $p_r$ (i.e., if more processing could be possible than there is still remaining in $c$, then the provider will have some non-utilised processing capabilities). Also, the possible processing is limited by the provider's maximum processing capability and the processing limit $p_l$ of the consumption.
\begin{eqnarray}
p_u^*(c,t+\SMALLESTIMEGRANULARITYM)&=&p_u(c,t)+\min\big(p_r(c,t),\\
&&\quad,\min(\mathfrak{p}(c, prov(c,t),t),p_l(c,t))\big) \label{EQ-PROV-UPDATE}\nonumber
\end{eqnarray}
Where $p_u^*: \bar{C} \times \SIMULATIONTIMEM \to \mathbb{R}^+$, $p_r: \bar{C} \times \SIMULATIONTIMEM \to \mathbb{R}^+$ and $p_l: \bar{C} \times \SIMULATIONTIMEM \to \mathbb{R}^+$ represent the processing under way, the remaining processing and the limit, respectively, for resource consumption $c$ at the time instance $t$. It must be pointed out that $p_u^*$ is offering the provider side under processing value only. Finally, $\mathfrak{p}: \bar{C} \times \bar{S} \times \SIMULATIONTIMEM \to \mathbb{R}^+$ reveals the processing power of a resource spreader (in this current case the provider for resource consumption $c$: $prov(c,t)$) at the time instance $t$. 

DISSECT-CF simulates consumers with similar behaviour, so they remove utilised resources from $p_u$. Of course in this case the limit of utilisation is dependent on the consumer and the previously evaluated provider side possible processing value:
\begin{eqnarray}
p_u(c,t+\SMALLESTIMEGRANULARITYM)&=&\max\big(0,p_u^*(c,t+\SMALLESTIMEGRANULARITYM)-\nonumber\\
&&\quad-\min(\mathfrak{p}(c, cons(c,t),t),p_l(c,t))\big)\nonumber\\\label{EQ-CONS-UPDATE}
p_r(c,t+\SMALLESTIMEGRANULARITYM)&=&p_r(c,t)+p_u(c,t+\SMALLESTIMEGRANULARITYM)-p_u(c,t)
\end{eqnarray}
Thus, to determine the state of a particular resource consumption, DISSECT-CF first evaluates the provider side of resource consumptions, and then it processes the consumer side. After the simulator determined the $p_u$ value for a resource consumption, the remaining consumption $p_r(c,t+\SMALLESTIMEGRANULARITYM)$ can be determined as well by reducing with the increment of the $p_u$ value. This behaviour ensures that at the end of the consumer side processing both $p_u$ and $p_r$ will represent the resource consumption's state in the next simulated time instance ($t+\SMALLESTIMEGRANULARITYM$). 

In order to determine how much processing can be done on resource consumptions at a particular time instance $\mathfrak{p}(c,s,t)$, resource spreaders apply the lowest level schedulers in DISSECT-CF based simulations. These schedulers share the processing capacities of the resources amongst those resource consumptions that the spreaders are currently dealing with~($\mathcal{P}: \bar{S} \times \SIMULATIONTIMEM \to \powerset(\bar{C})$, where the notation of $\powerset$ is used to depict a power set). As simulation developers are expected to run simulations with thousands of resources and millions of resource consumptions, these low-level schedulers must be highly customisable and efficient. To enable simple customisability, DISSECT-CF provides efficient implementations for most common scheduling related tasks (e.g., resource consumption registration, de-registration, event generation for parties interested in resource consumption state), allowing providers of new schedulers to just focus on the scheduling logic that calculates the new $\mathfrak{p}(c,s,t)$ values.

\subsubsection{Influence groups}
\begin{figure}[tb]
\centering
\subfloat[Generic influence groups of consumers and producers]{\includegraphics[height=4cm]{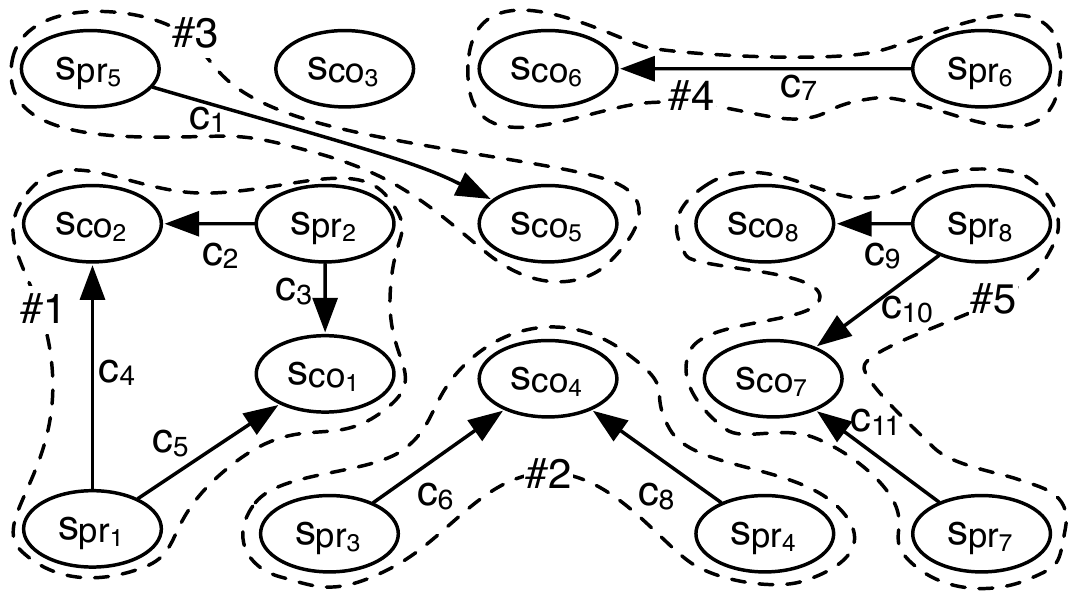}\label{FIG-INF-GEN}}\\
\subfloat[Influence groups formed from the CPUs of a physical machine]{\includegraphics[height=3.5cm]{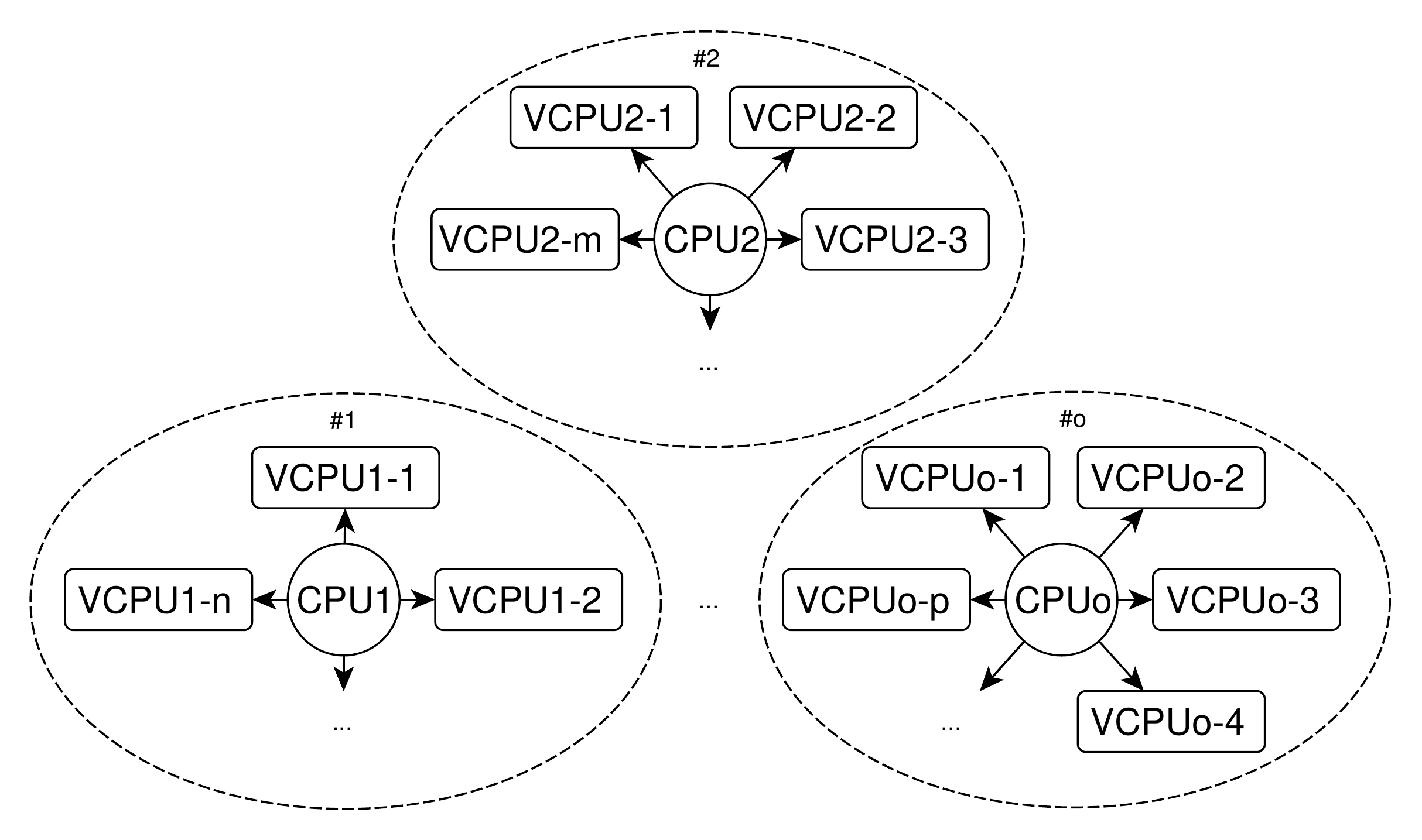}\label{FIG-INF-CPU}}
\caption{Examples of influence group formation\\
\emph{\footnotesize Notes: Edges represent resource consumptions and connect resource spreaders, specifically they point from providers to consumers. The identified influence groups are encircled with dashed lines. }\label{FIG-INFLUENCE}}
\end{figure}

The scheduling logic is expected to deliver fair resource allocation for simultaneously occurring resource consumptions -- denoted as $C: \SIMULATIONTIMEM \to \powerset(\bar{C})$. To simplify the behaviour and complexity of schedulers, DISSECT-CF also introduces the concept of influence groups. These groups are formed from all resource spreaders that have a chance to influence each other's resource allocation schedules. With the help of influence groups even schedules for complex network structures can be simulated at close-to-real-life behaviour (e.g., the simulator can apply fair share algorithms over multiple related network links). These schedulers can utilise influence groups as their domain in which they have to guarantee a fair resource schedule for the spreader associated resource consumptions.

To determine the membership of an influence group, the simulator uses the resource consumptions that link consumers and providers (see Figure~\ref{FIG-INF-GEN}). As a practical example, Figure~\ref{FIG-INF-CPU} shows how each simulated physical machine forms independent influence groups with the virtual machines it hosts via their respective CPU spreader implementations. Formally, an influence group of a resource spreader at the particular time instance is defined as follows ($G:  \bar{S} \times \SIMULATIONTIMEM\to \powerset(\bar{S})$):
\begin{equation}\small
G(s,t)=\{s\} \cup{\underset{s_o\in S(t):(\exists c\in \mathcal{P}(s,t):((prov(c,t)=s_o) \lor (cons(c,t)=s_o))}{\bigcup}G(s_o,t)}
\end{equation}
Where  $s \in \bar{S}$  is a resource spreader, and the function $S: \SIMULATIONTIMEM \to \powerset(\bar{S})$ defines the spreaders available at a particular time instance. The equation shows that $G(s,t)$ includes all resource spreaders that are directly or transitively referred by the associated resource consumptions of the spreader $s$~--~$\mathcal{P}(s,t)$.  As a result, one could find as many influence groups as the number of resource spreaders existing at a given time instance in the simulation. On the other hand, these groups are frequently equivalent because determining the influence group of any member of a particular group will result in the original influence group:
\begin{equation}
\nexists s_i\in G(s,t): (G(s,t)\backslash G(s_i,t)\neq\emptyset \land G(s_i,t)\backslash G(s,t)\neq\emptyset)
\end{equation}
This last equation is derived from the definition of $G(s,t)$ and reveals that after the proper calculation of $G(s,t)$ there should not be any members of it (e.g., $s_i$) that would result in a different influence group than the original $G(s,t)$.

Although the definition of $G$ is quite straightforward, its evaluation in all necessary time instances for all relevant resource spreaders would result in significant simulation performance deterioration. Therefore, DISSECT-CF provides an algorithm that significantly reduces the use of $G$ but still ensures that the influence groups are available for the scheduling logic in every time instance. To differentiate between the original function's results and the algorithm calculated values, the notation $G'$ is used for the influence groups determined by the algorithm (see Algorithm~\ref{ALG-BASOPT}).

\begin{algorithm}[tb]
\begin{algorithmic}[1]\small
\REQUIRE $G'(s,t)$ \COMMENT{Influence group in time instance $t$}
\STATE $G'(s,t+\SMALLESTIMEGRANULARITYM)\gets G'(s,t)$
\REPEAT \label{LIN-EXT-START}
\STATE $E\gets\emptyset$ \label{LIN-EMPTYEXT} \COMMENT{Future extension} 
\FORALL{$s_i\in G'(s,t+\SMALLESTIMEGRANULARITYM)$}
\STATE $add(s_i,t+\SMALLESTIMEGRANULARITYM) \gets \mathcal{P}(s_i,t+\SMALLESTIMEGRANULARITYM)\backslash \mathcal{P}(s_i,t)$ \label{LIN-IDENTADD}
\FORALL{$c: c\in add(s_i,t+\SMALLESTIMEGRANULARITYM) \land \big((prov(c,t+\SMALLESTIMEGRANULARITYM)\notin G'(s,t+\SMALLESTIMEGRANULARITYM)) \lor (cons(c,t+\SMALLESTIMEGRANULARITYM)\notin G'(s,t+\SMALLESTIMEGRANULARITYM))\big)$}
\STATE $E\gets \big(E \cup G'(prov(c,t+\SMALLESTIMEGRANULARITYM),t) \cup G'(cons(c,t+\SMALLESTIMEGRANULARITYM),t)\big)\backslash G'(s,t+\SMALLESTIMEGRANULARITYM)$ \label{LIN-EXTGADD}
\ENDFOR
\ENDFOR
\STATE $G'(s,t+\SMALLESTIMEGRANULARITYM)\gets G'(s,t+\SMALLESTIMEGRANULARITYM) \cup E$ \COMMENT{Actual extension} \label{LIN-REALEXT}
\FORALL{$s_i\in E$}
\STATE $G'(s_i,t+\SMALLESTIMEGRANULARITYM)\gets G'(s,t+\SMALLESTIMEGRANULARITYM)$
\ENDFOR \label{LIN-IGEXTCOMPLETE}
\UNTIL{$E=\emptyset$} \label{LIN-EXT-STOP}
\IF{$\exists s_x\in G'(s,t+\SMALLESTIMEGRANULARITYM): \big((\mathcal{P}(s_i,t)\backslash \mathcal{P}(s_i,t+\SMALLESTIMEGRANULARITYM))\neq\emptyset\big)$} \label{LIN-SPLIT-START}
\STATE $G_{temp}'\gets G'(s,t+\SMALLESTIMEGRANULARITYM)$
\REPEAT
\STATE $s_i \gets s\in G_{temp}'$ \COMMENT{random choice} \label{LIN-RANDSEL}
\STATE $G_{new}'\gets G(s_i,t+\SMALLESTIMEGRANULARITYM)$ \COMMENT{use of the original function $G$} \label{LIN-CORRECT-GROUP}
\STATE $G_{temp}'\gets G_{temp}'\backslash G_{new}'$ \COMMENT{Influence group splitting}
\FORALL{$s_j\in G_{new}'$}\label{LIN-SPLIT-UPD-START}
\STATE $G'(s_j,t+\SMALLESTIMEGRANULARITYM)\gets G_{new}'$
\ENDFOR\label{LIN-SPLIT-UPD-STOP}
\UNTIL{$G_{temp}'\neq\emptyset$}
\ENDIF \label{LIN-SPLIT-STOP}
\end{algorithmic}
\caption{Influence group management\label{ALG-BASOPT}}
\end{algorithm}

In the following few paragraphs, the internal behaviour of the new algorithm is discussed. It is built on the assumption that $\forall s\in S(0):G'(0,s)=s$ and it is composed of two distinct phases: influence group extension -- see lines \ref{LIN-EXT-START}-\ref{LIN-EXT-STOP} -- and group dissolution -- see lines \ref{LIN-SPLIT-START}-\ref{LIN-SPLIT-STOP}.

Let us first discuss the extension phase. During this phase, the algorithm first starts with an empty resource spreader set (see line \ref{LIN-EMPTYEXT}) that will later on hold the identified extensions of the input influence group -- $G'(s,t)$. As a next step, line \ref{LIN-IDENTADD} determines the resource consumptions that arrived to a particular resource spreader at the time instance $t+\SMALLESTIMEGRANULARITYM$. Afterwards, the following line focuses on those newly added resource consumptions that introduce new resource spreader members into the input influence group. These non-member providers or consumers are added to the extension set in line~\ref{LIN-EXTGADD}. This iteratively created extension set is used to actualise the input influence group in lines \ref{LIN-REALEXT}-\ref{LIN-IGEXTCOMPLETE}. The extension phase completes only when there are no newly introduced resource spreader members in the input group. Otherwise, the current phase is repeated to ensure finding even further group extensions via the resource consumptions associated with the introduced members (this last step is shown in line \ref{LIN-EXT-STOP}).

After there are no new extension possibilities found in the current influence group, the algorithm proceeds to its second phase in which it identifies all splits of the current influence group. For example, influence group \#5 in Figure~\ref{FIG-INF-GEN} will have to be split when the resource consumption between provider $\mathtt{s_{pr_8}}$ and consumer $\mathtt{s_{co_7}}$ finishes. To identify the need for splitting, the algorithm therefore first determines if there were some resource consumptions dropped from at least one member of the influence group (see line~\ref{LIN-SPLIT-START}). If there is a need for splitting, then the algorithm will maintain the not yet split parts of the original influence group in $G_{temp}'$. In order to determine which parts have to be split from the not yet split parts, lines \ref{LIN-RANDSEL} and \ref{LIN-CORRECT-GROUP} use the original $G(s,t+\SMALLESTIMEGRANULARITYM)$ on a randomly selected spreader from $G_{temp}'$ -- resulting in a new influence group called $G_{new}'$. In the next line, the not yet split parts are updated so that only those resource spreaders will be considered afterwards that are not in $G_{new}'$. Finally, the algorithm updates its self-maintained influence group membership so all members of $G_{new}'$ will be exactly the same (see lines \ref{LIN-SPLIT-UPD-START}-\ref{LIN-SPLIT-UPD-STOP}).

To conclude, the newly introduced algorithm reduces the number of direct $G(s,t)$ function calculations to those cases where there is a chance to have a group to be split. And even in that case, it ensures that the number of $G(s,t)$ evaluations is limited by the number of influence groups created after a split.

\subsubsection{Integration of low-level scheduling logic}

\begin{figure}[tb]
\centering
\includegraphics[width=\columnwidth]{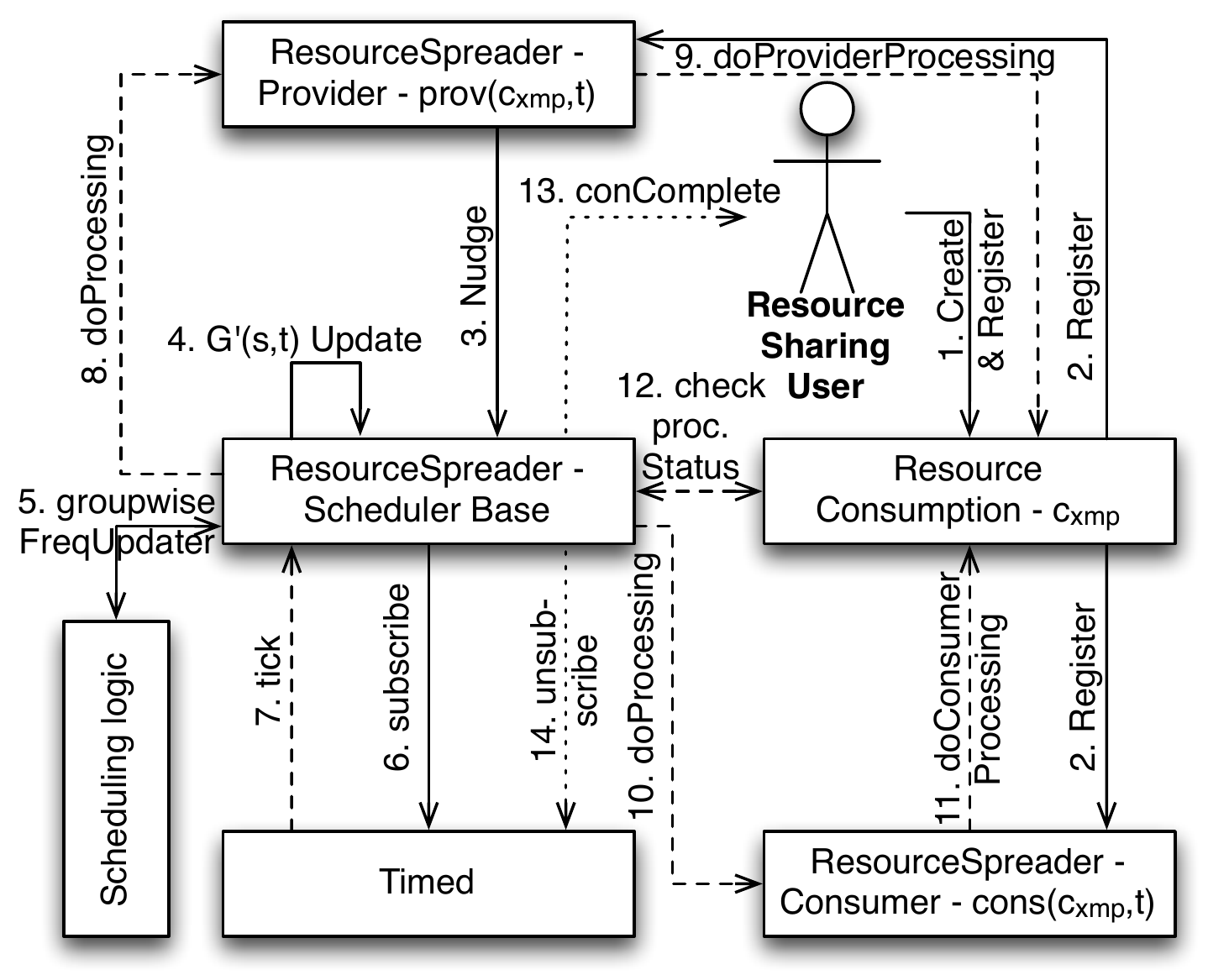}
\caption{The unified resource sharing mechanism and its relation to the scheduling logic provided by the simulation developer\label{FIG-RS}}
\end{figure}

The last remaining part of the unified resource model is the simulation developer customisable low-level scheduling logic. To understand the customisation options DISSECT-CF offers, Figure~\ref{FIG-RS} presents the context of the scheduling logic. The figure reveals the role of the low-level scheduler through the illustration of the life of a single resource consumption that can be represented in three phases and denoted with different kinds of arrows: preparation~--~regular lines; resource consumption~--~dashed lines; and completion~--~dotted lines. The next paragraphs provide a brief overview of these three phases.

First, the preparation phase is initiated by the entity who is responsible for creating (see \emph{Step 1} in the figure) a particular resource consumption -- $c_{xmp}$. This entity could be an automated process (e.g., a workload generator) or some higher-level entity of the simulator (e.g., the virtual machine representation). After creation, the registration can be initiated in any time instance $t$ after both the consumer -- $cons(c_{xmp},t)$ -- and the provider -- $prov(c_{xmp},t)$ -- spreaders are specified. The registration is accomplished in \emph{Step 2} in the figure. The provider nudges the scheduler base in \emph{Step 3} after both the consumer and the provider have registered the new resource consumption -- $c_{xmp}\in \mathcal{P}(prov(c_{xmp},t),t)$ and $c_{xmp}\in \mathcal{P}(cons(c_{xmp},t),t)$. The \emph{scheduler base} is implemented in the base class of all resource spreaders and is responsible for interfacing with the event system, the scheduler and the influence group management algorithm. Before contacting the event system for subscription, the scheduler base first updates the influence groups with Algorithm~\ref{ALG-BASOPT} in \emph{Step 4}. After the identification of all distinct influence groups, the scheduler base filters those groups that would need an updated schedule. Such groups are identified via recently added or dropped resource consumptions to/from one of their member resource spreaders -- i.e., $\forall s\in \bar{S}: \big(\mathcal{P}(s,t+\SMALLESTIMEGRANULARITYM) \backslash \mathcal{P}(s,t)\neq\emptyset\big)\lor\big(\mathcal{P}(s,t)\backslash \mathcal{P}(s,t+\SMALLESTIMEGRANULARITYM))\neq\emptyset\big)$. In \emph{Step 5}, the scheduling logic is invoked for each of the filtered groups. During this step, it should assign the $\mathfrak{p}(c,s,t)$ values for all resource consumptions that are currently taking place in a given influence group. With these assignments, the simulator calculates the earliest completion time of the currently managed resource consumptions. Then, in \emph{Step 6}, it subscribes to a notification from the event system in order to know when the next resource consumption will be removed from the filtered influence groups.

In the following phase, the simulator handles the resource consumptions. This phase is either done when the event system delivers the notification on a resource consumption completion (see \emph{Step 7}) or alternatively upon the registration of a new resource consumption. In the second case, the resource consumption handling is automatically executed before influence groups are calculated (i.e., Steps \emph{8-12} could precede \emph{Step 4} of the preparation phase if a resource consumption is registered at a resource spreader that has already had some prior resource consumptions). In practice, \emph{Steps 8-9} evaluate Eq. \ref{EQ-PROV-UPDATE} and \emph{Steps 10-11} evaluate Eq. \ref{EQ-CONS-UPDATE} for all simultaneously existing resource consumptions -- $c_x\in C(t)$ -- in the simulator. Resource consumptions -- e.g., $c\in C(t)$ -- are automatically marked for removal when they reach their completion -- i.e., $p_u(c,t)=0 \land p_r(c,t)=0$. As a final step for resource consumption handling, the scheduler base checks for resource consumptions marked for removal~(see \emph{Step 12}) and on all marked resource consumptions it executes the completion phase. 

In the final phase, the resource consumption's completion is simulated. Consumptions can be considered complete in two cases: either they have no further processing to be done or they were cancelled by the entity using the resource sharing mechanism. The scheduler base notifies this entity in both cases  -- see \emph{Step 13}. Then it checks if it has finished all the current resource consumptions -- $C(t)=\emptyset$. If there are still further resource consumptions to process, then the scheduler base resumes operations from \emph{Step 4}, otherwise it cancels further notifications from the event system.

As can be seen, the simulator expects scheduling logic implementations to utilise a fairly narrow and well-defined interface with the scheduler base. Through this interface the simulator ensures that whenever a new schedule is needed (i.e., new $\mathfrak{p}(c,s,t)$ values), the simulation developer provided scheduling logic is always called. DISSECT-CF also provides two sample implementations for this scheduling logic: a max min fairness algorithm~\cite{MaxMinFair} implementation with progressive filling, and a simple logic that does not deal with complex bottleneck situations but demonstrates the interfaces with the scheduler base.

\subsection{Energy Modelling} \label{Sec-EnergyModel}

Compared to other recently developed simulators, DISSECT-CF completely decouples energy modelling from resource simulation in order to allow accounting for such energy consumptions that are not in direct relation to the resource utilisation of data centres. With this approach, a more comprehensive energy and power modelling is achievable that enables the analysis of new sophisticated energy aware algorithms in the areas of virtual machine placement, task scheduling, etc. These algorithms previously were frequently limited because energy readings from heating, ventilation and air conditioning (HVAC) units or higher-level IaaS components (like VM schedulers or IaaS interfaces) were scarcely available in past simulators. Thus, this section presents how energy consumption related information is collected and accumulated so they can support future algorithms. Later, Section~\ref{Sec-HLS} discusses the foundations for physical machine schedulers that are expected to be the primary users of the models overviewed in this section.

First, in order to enable the decoupling, DISSECT-CF offers \emph{resource utilisation counters} both for producers and consumers. These counters allow an aggregated and time dependent view of the consumption of particular resources. Counters are updated depending on the \verb+power state+ of a resource spreader (see Figure~\ref{FIG-ARCH}). For example, a physical machine -- $s_{pm} \in S(t)$ -- in suspend to RAM (STR) power state zeroes its processing power: 
\begin{equation}
\mathfrak{p}(c,s_{pm},t):=0\quad\mathrm{for}\quad\forall c\in C(t):prov(c)=s_{pm}
\end{equation}
The power states of the various entities in the simulation can be defined as needed; the simulator only expects these states to define the basic power characteristics (e.g., minimum and maximum power draw) and the resource processing behaviour of the given entity at the specific state. The simulator also provides a basic set of power states (on, off, turning on, turning off) for which the resource processing behaviour is already defined for the resources incorporated into a physical machine.

Based on this low-level power modelling functionality, the decoupling of energy models is accomplished through  \verb+energy meter+s. These meters are organised around four functionalities: $(i)$ monitoring of energy consumption directly related to resource utilisation, $(ii)$ indirect energy consumption estimation, $(iii)$ aggregation of metering results from multiple meters and $(iv)$ presenting up-to-date energy readings to their users. The remainder of this subsection reveals how these functions accomplish the decoupling and shows the ways customised, infrastructure specific metering can be achieved in DISSECT-CF.

\subsubsection{Energy metering} \label{SEC-ENMET}
\begin{figure}[tb]
\centering
\includegraphics[width=\columnwidth]{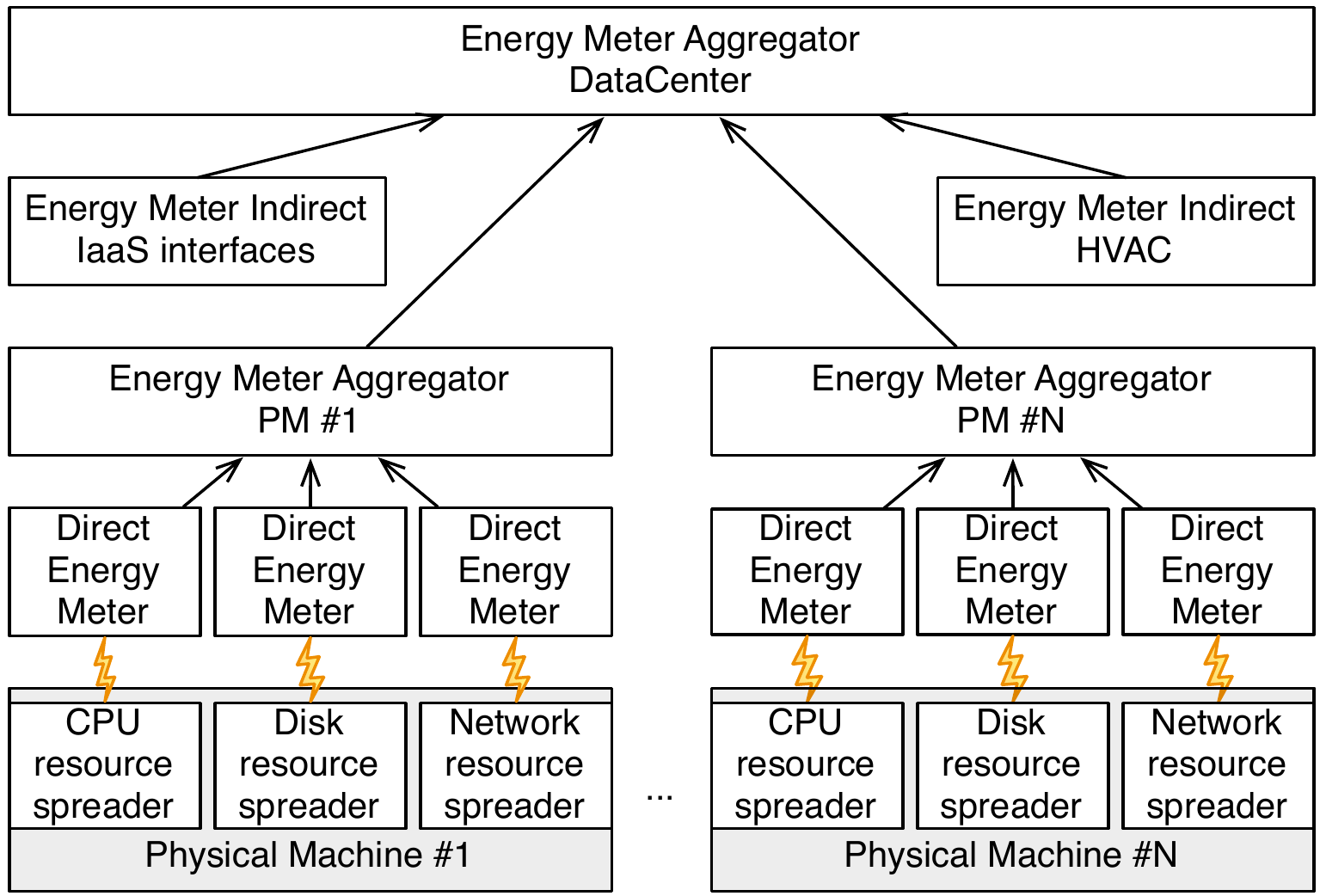}
\caption{Hierarchical metering in DISSECT-CF\label{FIG-METERS}}
\end{figure}

\paragraph{Direct resource utilisation related energy consumption metering} Based on the previously mentioned resource utilisation counters, the simulator can be requested to periodically evaluate the instantaneous utilisation percentage. Then, energy \verb+consumption model+s use these percentages (see Figure~\ref{FIG-ARCH}) to estimate the instantaneous power draw of each monitored resource spreader (later the resulting estimate and the metering period is used to calculate the direct meter's energy consumption estimate). Consumption models are dependent on the actual power state a resource spreader is in, and the simulator developer can define them. As examples, the simulator provides two simple energy consumption model implementations: $(i)$ a linear interpolation between minimum and maximum power draw depending on current resource utilisation, to allow basic modelling of  dynamic power behaviour -- or $(ii)$ a constant minimum power draw, to allow the effortless modelling of \emph{off} or \emph{STR} power states. Figure~\ref{FIG-METERS} presents direct meters for each resource spreader in the shown physical machines.

\paragraph{Indirect energy consumption estimation} To support more complex energy consumption estimates, the simulator also allows energy consumption to be derived from other properties of the simulated system. For example, these properties could include the virtual machine request rate of a particular data centre, or the utilisation of the data centre level storage subsystem (e.g., to estimate how many disk drives the currently stored data can occupy). These meters are expected to periodically evaluate the system state and accumulate their energy consumption estimates for those components of the simulated system that are not directly represented with resource spreaders. Figure~\ref{FIG-METERS} reveals two indirect metering solutions to represent the internal actions of IaaS systems and the energy consumption behaviour of data centre level HVAC systems.

\paragraph{Meter aggregators} In several cases, the energy consumption values from individual direct or indirect meters are not sufficient for higher-level energy aware decision makers (e.g., physical machine state schedulers -- see Section \ref{sec-schedulers}). For example, a physical machine in DISSECT-CF is represented with multiple resource spreaders (e.g., CPU, disk bandwidth), thus to have a complete view of a physical machine's energy consumption, one would need to monitor several direct energy meters. Meter aggregators allow the automated collection and management of several meters in parallel and provide the higher-level view expected by decision makers. Figure~\ref{FIG-METERS} shows a complex scenario for the use of aggregators. This scenario shows that a single aggregated meter can be constructed for a whole data centre, allowing even the inclusion of indirect metering results such as HVAC. The figure also shows how a physical machine's resource spreader set can be metered as a single entity.

\subsubsection{Energy consumption accounting strategies}

As the estimation of energy consumption of the simulated entities might be time consuming, the simulator allows simulation developers to fragment and focus their measurements on those parts of the simulated systems that they are interested in. For instance, the simulation developers might be only interested in the energy consumption of a single virtual machine that is deployed on a simulated cloud with multiple data centres. In such cases, DISSECT-CF can limit the number of energy meters that are evaluated with two approaches: independent meters or adjusted aggregations.

\paragraph{Independent meters} This approach entails that metering results are only dependent on the metered component and the rest of the simulated system cannot influence them. E.g., the energy consumption reported for a CPU level resource spreader of a physical machine should not be dependent on the behaviour of the rest of the system.

\paragraph{Adjusted aggregations} If there is a dependency between two metered components (e.g., a virtual machine that is hosted on a particular physical machine), then a special meter aggregator can be created. This meter aggregator will not just add the aggregated meters' measurements. Instead it allows simulation developers to define an aggregation function.

DISSECT-CF uses the adjusted aggregation technique to handle derived energy consumptions such as VM level energy consumption. For example, when applying the linear interpolation based energy consumption model, virtual machine level power draw can be calculated using the power draw of the hosting physical machine, as follows:
\begin{equation} \label{EQ-POWERMODEL}
P_{vm}=P_{pm}'\frac{\underset{\forall c\in C(t):cons(c,t)=s_{vm}}{\sum}\mathfrak{p}(c,s_{vm},t)}{\underset{\forall c\in C(t):prov(c,t)=s_{pm}}{\sum}\mathfrak{p}(c,s_{pm},t)}+\frac{P_{pm}^{idle}}{|G(s_{vm},t)|-1}
\end{equation}
Where $P_{vm}$ is the derived instantaneous power draw of a particular VM while it is running. The equation's first part estimates the variable part of the power draw, while the second part provides an estimate for the idle part. The variable part is dependent on $P_{pm}'$, which is the maximum variability of the physical machine's power draw. The variable part is proportional to $P_{pm}'$, depending on the resource utilisation of the particular VM compared to the resource utilisation of all VMs hosted on the same physical machine. The second part of the estimate is the idle part that is derived from the idle power draw of the physical machine -- $P_{pm}^{idle}$. This part is proportional to the number of VMs hosted by the physical machine at the given time instance (this number is one less than the cardinality of the VM's influence group because the group contains also the resource spreader of its hosting physical machine).

Eq. \ref{EQ-POWERMODEL} reveals that the energy consumption model (that estimates the instantaneous power draw) of the virtual machine is not independent from the physical machine's behaviour. Therefore, energy consumption cannot be directly accounted to virtual machines. When such meters are requested, the simulator identifies them as dependent meters and instead of creating independent meters, it creates an adjusted meter aggregator including those meters that could provide the necessary information to calculate the energy consumption to be attributed to the originally requested meter. It must be noted that dependent meters consider energy consumptions multiple times (e.g., energy consumption is accounted to both physical and virtual machines). Thus when meter aggregations are created they must only include meters that are not dependent on each other.

\subsection{Infrastructure Simulation}

To allow the simplified development of new VM placement algorithms and PM state schedulers, DISSECT-CF provides an implementation of relevant infrastructure components in IaaS systems. These components are built on top of the previously discussed resource sharing and energy modelling techniques and provide abstractions for networked entities and for physical/virtual machines (see Figure~\ref{FIG-ARCH}). 

\subsubsection{Networking} \label{Sec-NW}
Network activities rarely play a role in scheduling decisions related to tasks or to physical/virtual machines. Thus, to increase the performance of simulations, by default, DISSECT-CF offers a limited network model where two networked entities must be always directly connected (therefore connection properties like bandwidth must be defined between all networked entities that should be able to communicate with each other in the simulation). This rudimentary behaviour could be sufficient even for some network aware schedulers, but to allow better representation of real networks, the simulator also allows the creation of intermediary network entities (such as routers). The implementation of such entities should alter the processing limit ($p_l(c,t)$) of all resource consumptions that are directed through them.

Directly connected networked entities ($n\in \mathcal{N}$, where $\mathcal{N}$ is the set of all possible networking entities) are simulated with the \verb+NetworkNode+ component. This component encapsulates an incoming and an outgoing network connection simulated with the unified resource sharing foundation; thus connections are implemented as resource spreaders: $n=<s_{in},s_{out}>$, where $s_{in},s_{out}\in\bar{S}$. The processing power of these spreaders represents the network bandwidth (either incoming or outgoing) of the given network node. When a new network communication must take place, simulation developers are expected to request a resource consumption between the source network node's outgoing resource spreader and the target's incoming resource spreader. The component also introduces network latencies ($l:\mathcal{N}^2\times \SIMULATIONTIMEM\to\mathbb{N}$) that can be defined between every networked entity for any given time instance. The latency values resulting from this function are used as delays preceding the registration of each resource consumption to the incoming or outgoing resource spreader of the node. For example, let us see what happens if a network communication (represented as a resource consumption $c_{reg}\in\bar{C}$) needs to be registered between a source and a target networked entity ($n_{source}, n_{target}\in \mathcal{N}$)  at the time instance $t_{reg}\in\SIMULATIONTIMEM$:
\begin{eqnarray}
n_{source}&=&<s_{in_s},s_{out_s}>\\
n_{target}&=&<s_{in_t},s_{out_t}>\\
l_{reg}&=&l(n_{source},n_{target},t_{reg})\\
prov(c_{reg},t)&=&\left\{\begin{array}{ll}
s_{nil} & \mathrm{if}\quad t<t_{reg}+l_{reg}\\
s_{out_s} & \mathrm{otherwise}
\end{array}\right.
\\
cons(c_{reg},t)&=&\left\{\begin{array}{ll}
s_{nil} & \mathrm{if}\quad t<t_{reg}+l_{reg}\\
s_{in_t} & \mathrm{otherwise}
\end{array}\right.
\end{eqnarray}
Where $l_{reg}$ is the network latency between the source and target nodes at the time of registration ($t_{reg}$), and $s_{nil}\in\bar{S}$ is the nonperforming spreader that never processes any resource consumption: $\forall c\in\bar{C}, \forall t\in\SIMULATIONTIMEM:\mathfrak{p}(c,s_{nil},t)=0$. Thus the equation shows, that the consumption is registered to the nonperforming spreader for the complete period of the latency, afterwards the simulator switches the consumption's registration to the originally designated spreaders (which are the network output port $s_{out_s}$ of the source node $n_{source}$ and the input port $s_{in_t}$ of the target node $n_{target}$). This last step allows the simulator to utilise its unified resource sharing mechanism after the latency period is over.

\subsubsection{The behaviour of physical machines}

In IaaS systems, physical machines offer most of the user exploited simulated resources. Thus, DISSECT-CF \verb+physical+ \verb+machine+s encapsulate a diverse set of resources: local disks (via Repositories -- see Section~\ref{SEC-UI}), network interfaces (with the help of network nodes -- see Section~\ref{Sec-NW}), CPUs (using the unified resource sharing model of Section~\ref{Sec-URSM}) and memory. Besides the modelling of these resources, DISSECT-CF's physical machine behaviour also focuses on two additional functionalities: administering resource allocations and VM requests; and modelling physical machine level power behaviour. As resource sharing and modelling has already been discussed in detail, this subsection mainly discusses the latter functionalities.

\paragraph{Resource allocator and VM request handler} In order to maintain up-to-date information on the available and utilised resources of the physical machine, DISSECT-CF applies a resource allocator. The applied allocator can reserve resources (e.g., a given amount of memory or number of CPU cores) from the physical machine. These reservations are represented as \verb+resource+ \verb+allocation+ instances that are used to maintain the free resource set of the physical machine. Allocation instances are also passed as a token of resource availability towards virtual machine schedulers. Other than the amount of resources associated with them, resource allocations also have the following properties: expiry time if unused or a link to the VM that uses it. When a resource allocation is created, the physical machine automatically initiates a \verb+DeferredEvent+, which automatically cancels the allocation after the expiry time. To avoid this mechanism, the entity that requested the allocation must request a VM to use the reserved resources (i.e., establishing the link between the VM and the allocation). The automatic cancellation of the resource allocation is a self-defence mechanism of physical machines to avoid keeping resources out of use just because they received an unused allocation.

For a single VM request, DISSECT-CF allows multiple \verb+resource+ \verb+allocation+s to be made across multiple physical machines. The multi-allocation technique can be used by schedulers to optimise for non-functional properties (e.g., past availability, expected energy consumption, environmental impact) of those resources that a VM could bind to at a given moment. After a decision is made about the use of a particular allocation for the VM request, the rest of the allocations (which were non optimal according to the non-functional requirements) are expected to be cancelled by the schedulers. For complex VM instantiation scenarios, schedulers are also allowed to adjust the expiration time upon allocation request. With this mechanism, researchers can evaluate advanced reservation-like scenarios regarding VM instances.

\paragraph{Power behaviour} 

\begin{figure}[tb]
\centering
\includegraphics[width=\columnwidth, trim=3cm 6.8cm 7.1cm 3.2cm, clip=true]{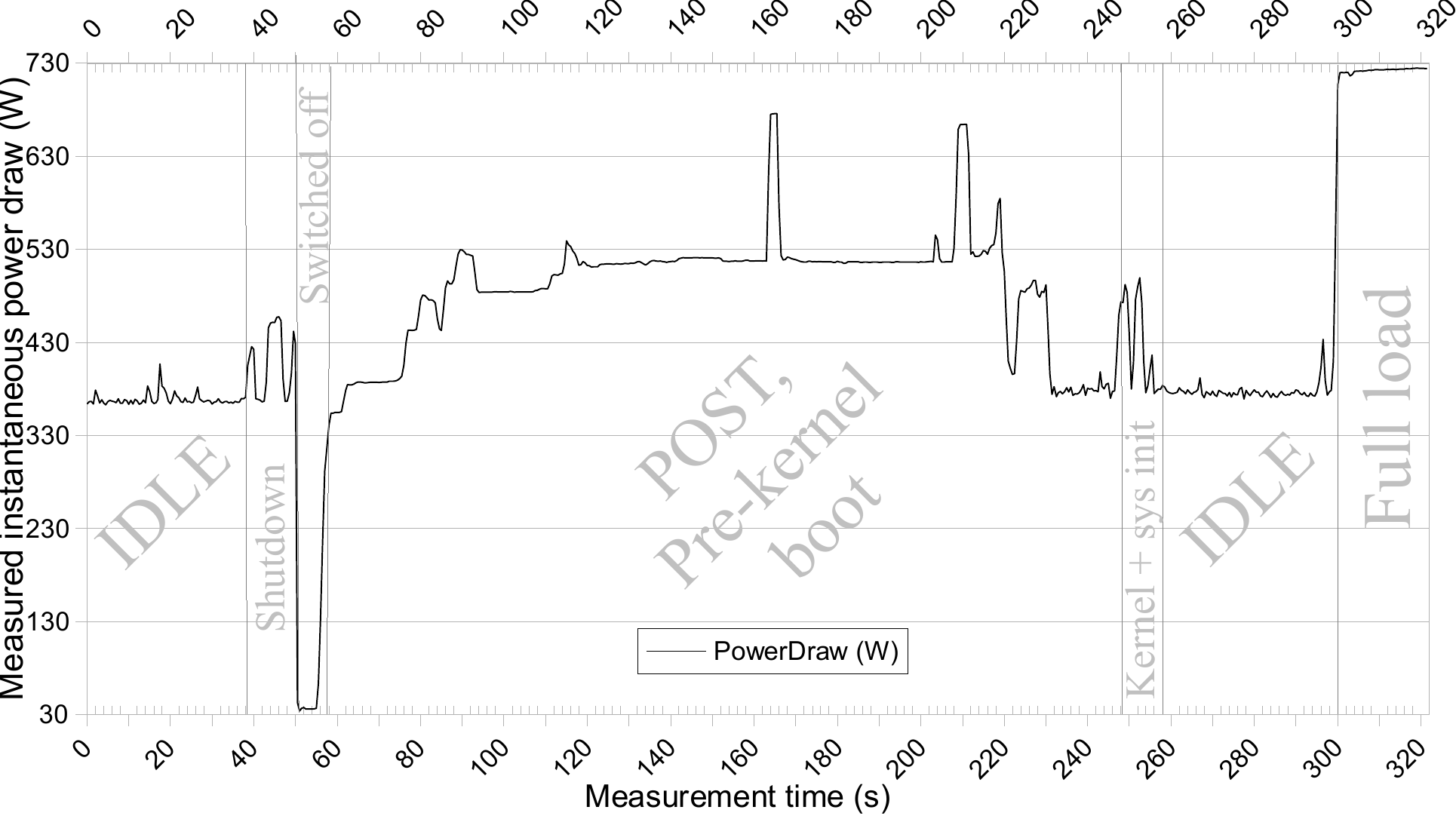}
\caption{Measured power behaviour of a cloud node\label{FIG-PMbehaviour}}
\end{figure}

As DISSECT-CF aims at supporting the development of energy aware scheduling strategies in IaaS clouds, the energy model of the physical machine is particularly important. In its default configuration, the simulator supports 4 power states (and the transitions between them): off, switching on, running and switching off. Although this power state set is fairly limited, the simulator already offers constructs that allow the modelling of more complex operations like: suspend to RAM, suspend to disk, dynamic voltage \& frequency scaling or core/CPU de- and reactivation. Suspension related states can be modelled with the introduction of new power states, while the latter two are available because resource spreaders can alter their maximum processing capabilities.

The modelling of the 4 supported states was done after real-life physical machine behaviour in the clouds of the University of Innsbruck and MTA SZTAKI. The real-life behaviour of the machines was observed through constantly monitoring their instantaneous power draw while they went through the following cycle: idling $\to$ shutdown $\to$ off $\to$ switch on $\to$ idling $\to$ full CPU load. A sample measurement with a typical cloud node at Innsbruck (with 80 CPU cores, 128GB SSD, 132GB RAM, and redundant power supplies) can be seen in Figure~\ref{FIG-PMbehaviour}. As physical machine state schedulers can be highly influenced by the behaviour in non-running states (e.g., their decision on shutting off a machine could be dependent on the time it takes to boot the machine back and the expected power savings because of the completely off machine), DISSECT-CF offers a simplified and a more complex behaviour model. 

\begin{table}[tb]
\centering
\caption{Simplified power state definition of a physical machine\label{TAB-SIMPLI}}
\begin{tabular}{p{2cm}p{1.2cm}p{1.3cm}p{1.3cm}p{0.9cm}}
Power state & Cons. model & Min cons. & Max cons. & Dura-tion \\
\hline
Off & Constant & 36.4 W & -- & N/A\\
Switching on & Constant &483.1 W & -- & 200 s\\
Running & Linear & 368.8 W & 722.7 W & N/A\\
Switching off & Constant & 409.2 W & -- & 12 s\\
\hline
\end{tabular}
\end{table}

In the simplified model, all states except the running state are modelled with the constant minimum power draw technique (see Section~\ref{Sec-EnergyModel}). Thus when defining the power states for the machine, simulation developers are expected to provide only the average power draw and state duration figures. Thus, in the simplified model the machine measured in Figure~\ref{FIG-PMbehaviour} can be defined with the power states listed in Table~\ref{TAB-SIMPLI}.

On the other hand, the complex behaviour introduces a hidden consumer to the physical machine's resources. This consumer is able to consume resources in all power states except the off state, allowing a more fine-grained approximation of resource utilisation; therefore, it is closer to real-life energy characteristics of physical machines. The hidden consumer is also important to model resource consumptions for virtual machine related activities done by the virtual machine monitor or hypervisor (e.g., creation, migration). If the simulation developers would like to define complex physical machine behaviour then they are expected to provide a set of tasks (in the form of CPU, network and disk consumptions) for the hidden consumer in every power state. The simulator will take these tasks and will execute them in the order and timing specified by the simulation developer. The simulator assumes that completing all the specified tasks marks the end of a particular power state. Thus, in the complex behaviour model, simulation developers should define the machine from Figure~\ref{FIG-PMbehaviour} according to Table~\ref{TAB-COMPLEX} and they should also define the hidden consumer's tasks for the switching on/off states. For example, for switching off they could define the following resource consumptions (assuming that $\SMALLESTIMEGRANULARITYM=1s$):
\begin{eqnarray}
c_1&=&<0, 0.275, 0.11> \label{EQ-C1}\\
c_2&=&<0,0.855,0.19>\\
c_3&=&<0,0.228,0.15>\label{EQ-C3}
\end{eqnarray}
Where $c_1$ is registered to the hidden consumer at the start of the shutdown state, then $c_2$ is 2.5 seconds after the completion of resource consumption $c_1$, and $c_3$ is registered one second after $c_2$ is completed (the registration timings must be independently specified from the resource consumption definitions, therefore Equations \ref{EQ-C1}-\ref{EQ-C3} do not contain references to these timings). For simplicity, in the example, each resource consumption's $p_l$ value (e.g., $p_l(c_1)=0.11$) is specified assuming that the physical machine's total processing capabilities in a second is represented with one (thus $c_1$'s value of 0.11 means it can maximally utilise 11\% of the processing capabilities of the machine).

\begin{table}[tb]
\centering
\caption{Complex power state definition of a physical machine \\
{\footnotesize \emph{Remark:} in the switching on/off states the complex definition must also include tasks for the hidden consumer}
\label{TAB-COMPLEX}}
\begin{tabular}{ccrr}
Power state & Cons. model & Min cons. & Max cons. \\
\hline
Off & Constant & 36.4 W & -- \\
Switching on & Linear &368.8 W & 722.7 W\\
Running & Linear & 368.8 W & 722.7 W\\
Switching off & Linear & 368.8 W & 722.7 W\\
\hline
\end{tabular}
\end{table}

\subsubsection{Virtual machine model}

The DISSECT-CF model of virtual machines focuses on two main aspects of virtual machine behaviour: $(i)$ resource sharing for tasks executed by IaaS users, and $(ii)$ state management and transitions to support virtual machine placement and pm state scheduling techniques. The following paragraphs discuss these two aspects in detail.

\paragraph{Resource sharing} Virtual machines are the primary resource consumers in the DISSECT-CF based simulations. They receive the user's tasks and transform them to various resource consumptions (this functionality is accomplished in the virtual machine's \verb+newTask+ function). The simulator primarily focuses on CPU based resource consumptions (by creating a \verb+resource+ \verb+spreader+ for CPU consumptions); however, simulation developers can also request network consumptions. As memory and disk activities for tasks cannot be monitored well, DISSECT-CF does not expect simulation developers to be able to describe their application's behaviour with regards to these two activities. For memory, the simulator simply ensures that the allocated memory of a VM will not be accessible by any other VM on the host. For disk activities, the simulator only focuses on transferring and accessing the disk image of the virtual machine. Disk image access is only simulated when the disk image is stored remotely; thus, disk access must be transformed to network activities.

Some simulators provide a representation for the network between the physical machine and its hosted virtual machines. This network, however, behaves very differently from regular networking because of the virtual machine monitor's (e.g., Xen, kvm) behaviour. The communication between virtual machines hosted on the same physical machine is more comparable to local inter-process communication mechanisms and thus it is practically instantaneous and has very low latencies. DISSECT-CF therefore does not offer networks between virtual machines hosted on the same physical machine, and in fact it does not allow VMs to act as networked entities. Instead, when a VM needs to communicate over the network, it has to use the physical machine's connections as the source or target of its communication.

\paragraph{State management} The following list describes the behaviour of the virtual machine in the various states represented in Figure~\ref{FIG-VMStates}:
\begin{figure}[tb]
\centering
\includegraphics[width=\columnwidth]{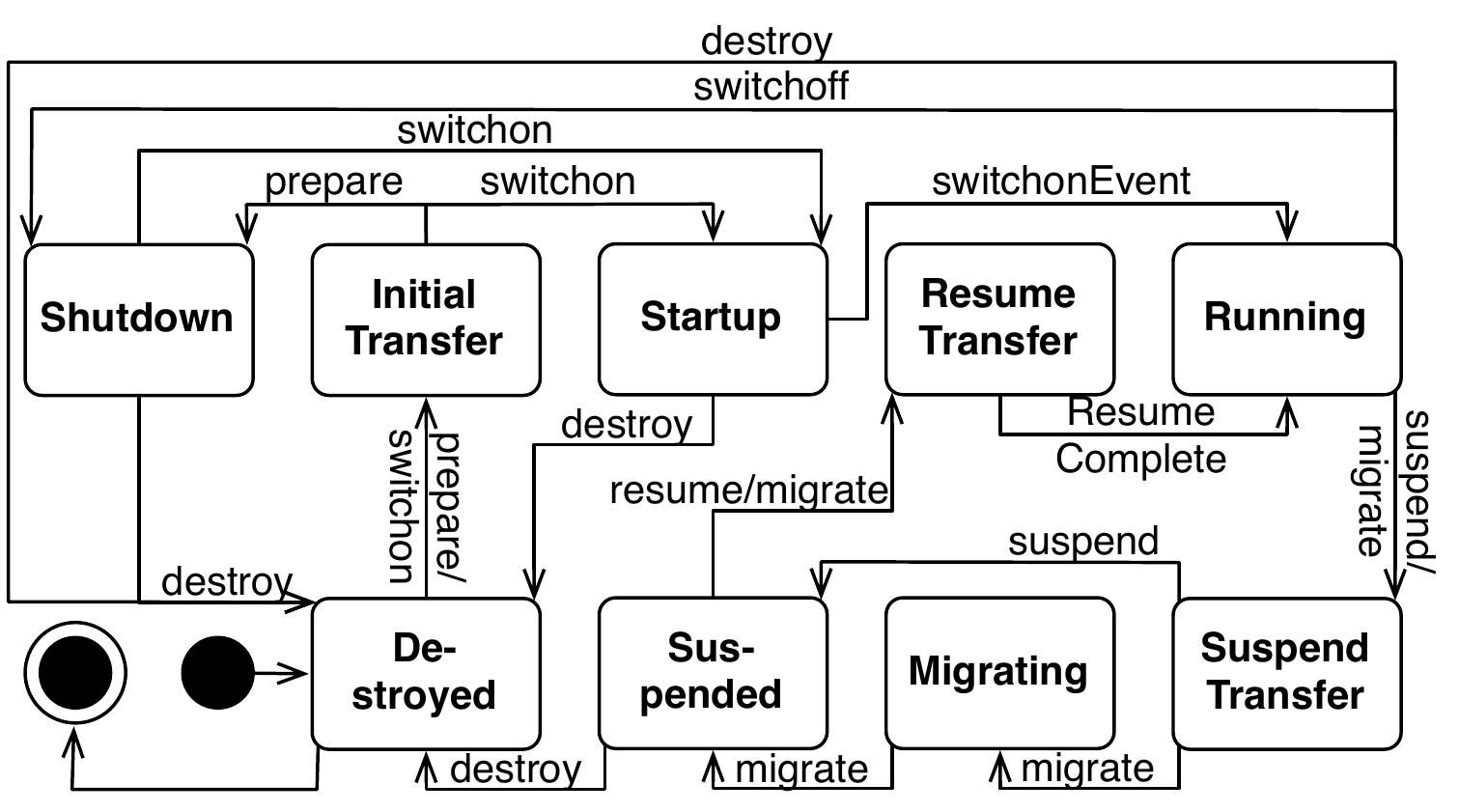}
\caption{Virtual Machine state diagram\label{FIG-VMStates}}
\end{figure}

\begin{description*}
\item[Destroyed.] This is the initial state of every VM in DISSECT-CF. During this state, the VM is not consuming any kind of resources but it already allows monitoring of its state.
\item[Initial transfer.] This state shows that the image for the virtual machine is being transferred to its hosting location. This process creates a unique copy of a virtual machine image that represents the VM's functionality. Depending on the internal organisation of the cloud infrastructure, the hosting location of the image could be either the local disk of the physical machine that will host the VM or alternatively a central highly accessible storage system.
\item[Shutdown.] When the image of the VM is at its hosting location and it is not occupying any other resources, then it is in this state. Allowing for a separate shutdown state enables two  techniques: $(i)$ image pre-staging -- when a VM placement algorithm prepares an image in advance to an actual VM request --, and $(ii)$ disk state preservation -- when an IaaS user does not intend to use its VM for a while but he/she would like to be able to continue his/her work from the same disk state as before the shutdown. 
\item[Startup.] This state represents the boot-up procedure of the VM. When this state is reached, the VM is bound to a physical machine's \verb+resource+ \verb+allocation+, and thus the VM is already capable of consuming some resources from the physical machine. When a virtual machine image is defined, simulation developers must specify the resource requirements for the boot-up procedure of a particular VM functionality. During start-up, these resource requirements are transformed into resource consumptions from the physical machine's resources. When all the transformed resource consumptions are complete, the VM is considered booted up and ready for executing user tasks.
\item[Running.] In this state, the VM is awaiting user requests for new tasks, as described in the resource sharing related paragraphs above.
\item[Suspend transfer.] Through this state, DISSECT-CF simulates the serialisation of the  VM's memory state. At the current implementation this is done when the VM has stopped its resource consumptions. To allow serialisation while the VM is still processing resource consumptions, the simulator would need a more comprehensive memory model where one cannot only see the memory utilisation in terms of bandwidth and page hits but also in terms of page level access patterns. As this model would require very specific application descriptions and would result in longer runtimes, it is not considered in the current simulator.
\item[Migrating.] This state shows when the VM's serialised state is transferred between two physical machines. After the transfer, the VM can be resumed on a different physical machine than the one that it was originally hosted on. Thus, migration plays a key role in server consolidation techniques applied by PM state schedulers. Also, through the simulated migration solution of DISSECT-CF, VM schedulers will be able to rearrange server load, allowing prioritisation of highly utilised VMs while still ensuring some resources for under-utilised ones.
\item[Suspended.] This state is similar to the shutdown state, except that alongside the VM's image the VM's memory state is also stored.
\item[Resume transfer.] During this state, the reloading of the VM's memory state is simulated. At the end of the state the VM will again become ready for executing tasks. Those tasks that were already in the process of execution when the suspend or migration operation was requested are also restored and their resource consumptions continue from the state before their suspension.
\end{description*}

\subsection{Infrastructure Management} \label{Sec-IM}

As seen in Figure~\ref{FIG-ARCH}, the top-level components of the simulator are the VM scheduler, the PM scheduler, the Repository, and the IaaS service. Although these components are sitting on top of the rest of DISSECT-CF, only the Repository and the IaaS service components are directly accessible to users. Thus, first the section discusses how the PM and VM schedulers organise the internals of a single infrastructure cloud, then the section concludes with the expected use cases of the remaining components.

\subsubsection{The high level schedulers} \label{Sec-HLS}

Similarly to the unified resource-sharing model, the two high level schedulers in DISSECT-CF also contain just a high performance foundation of custom schedulers provided by the simulation developers. In case of \verb+VM+ \verb+scheduler+s this foundation is responsible for $(i)$ receiving VM requests; $(ii)$ analysing their feasibility -- i.e., if they could ever be hosted on the underlying physical machines of the cloud; $(iii)$ managing request queues, $(iv)$ keeping in contact with PM schedulers so they know the state of the request queue; and finally $(v)$ dispatching scheduling requests to the custom schedulers when there is a chance for a VM to be placed on the infrastructure. Custom schedulers therefore are expected only to contact physical machines for resource allocations, associating VMs with the allocations and removing served requests from the queue.

DISSECT-CF provides three custom scheduler implementations that are delivered with the simulator and show how a VM scheduler can be implemented. These schedulers implement the first fit algorithm in three different approaches: $(i)$ basic -- the earliest not yet served virtual machine request is attempted to be placed on any available physical machine; if no machine can host the request, then no further requests are processed; $(ii)$ non-queuing  -- always ensures the rejection of those requests that cannot fit into the infrastructure currently; and $(iii)$ minimal request first -- orders the queue so it contains the smaller-sized requests first, but otherwise runs like the basic approach.

In the case of \verb+PM+ \verb+scheduler+s, the foundation is responsible for detecting VM request queuing and infrastructure size changes (i.e., the removal or addition of a physical machine). The actual scheduling code is expected to handle the queuing and infrastructure size changes in order to maintain a physical machine set that is capable of promptly serving upcoming VM requests but utilises as little energy as possible. PM schedulers are expected to optimise the load of physical machines by methods such as automatically consolidating servers through the migration of dormant virtual machines. Finally, PM Schedulers could also utilise the energy readings of the controlled physical machines in order to optimise the overall energy behaviour of a particular data centre.

The simulator is delivered with two PM state schedulers: one that keeps all physical machines in running state and one that schedules PMs depending on the VM schedule. The latter only turns on new machines when there is a growing queue of VM requests that cannot be served by currently running machines. This scheduler is also able to turn off machines if the VM queue is empty and there are some physical machines without any load.

\subsubsection{External IaaS APIs} \label{SEC-UI}

So far only the internals of the simulator have been discussed. Now let us turn our attention to the external APIs that enable the investigation of user-side schedulers (i.e., Task to VM assignment, Virtual Infrastructure scaling or Cross-cloud VM allocation -- see Section~\ref{sec-schedulers} for details). These external APIs have three functionalities: $(i)$ serve as the primary information source about a particular cloud infrastructure; $(ii)$ allow the management of virtual infrastructure components; and $(iii)$ offer infrastructure alteration capabilities. 

\paragraph{APIs for information retrieval} These APIs offer information for the decision-making process of the user-side schedulers. Through the \verb+IaaSService+ component, one can collect information about:
\begin{itemize}
\item the ratio of running and total number of physical machines (e.g., $\frac{1}{3}$ of the infrastructure is still not turned on);
\item the number of hosted virtual machines (allows the user-side scheduler to easily grasp the current load of the system without going into details);
\item the total and running capacity (in terms of offered resources by physical machines -- e.g.,  500 CPUs total, 200 are already running);
\item the list of physical machines (e.g., to access more fine-grained system utilisation details such as the load of each running machine);
\item the currently applied PM and VM schedulers (e.g., this allows user-side schedulers to send specific requirements with their VM requests according to the applied VM schedulers).
\end{itemize}
One should bear in mind that this kind of information is at best available only partially in current commercial and  academic cloud(ware) offerings -- e.g., Amazon EC2\footnote{http://aws.amazon.com/ec2}, OpenNebula~\cite{opennebula2011}. Therefore, nowadays, user-side schedulers must predict the necessary information based on what cloud providers offer (e.g., turning the spot pricing information of Amazon to a similar measure as the ratio of running and total number of PMs). However, the newly available information in DISSECT-CF allows the analysis of user-side schedulers from new perspectives (i.e., what difference does it make if a scheduler utilises some of the new information). Based on the results of this analysis, IaaS providers will be able to offer the most useful information for such schedulers in the future.

The \verb+Repository+ component of the simulator provides additional support for user-side schedulers regarding data and virtual machine image storage: $(i)$ the set of repositories of an infrastructure can be queried; $(ii)$ it is possible to search for contents (e.g., to choose a VM hosting infrastructure that has relevant data stored locally); $(iii)$ one can determine the size of unused storage capacities (e.g., to allow checking if cross-cloud migration of data would be suitable); and finally $(iv)$ the available bandwidth of the disk and network interfaces of each repository can be queried to determine how fast it can send/receive the required contents for a particular virtual infrastructure.

\paragraph{APIs for virtual infrastructure management} Once a decision is made about VMs, user-side schedulers can contact the \verb+IaaSService+ for the following activities: VM request, VM termination, VM resource reallocation (allowing down or up-scaling VMs while they are running) and cross-cloud VM migration (enables users to leave a particular cloud but continue their activities in their current VMs in some other cloud). Concerning the \verb+Repository+ component, user-side schedulers can request data to be transferred between repositories within or even across-cloud providers.

\paragraph{APIs for infrastructure alteration} In the initial phases of the simulation, one needs to construct the clouds that the simulation will use as its infrastructure. This API allows the registration and de-registration of physical machines and repositories. Inside repositories it also allows instantaneous registration or de-registration of stored data or virtual machine images (e.g., as if the data was already there on the repository without the need for its actual transfer). Naturally, user-side schedulers do not interact with these APIs directly. However, in some cases, they should react to the evolution of the infrastructure utilised during their operation. Infrastructure alteration capabilities thus enable the evaluation of these schedulers in a dynamic context (e.g., checking for error resilience or autonomous operations). For example, when a physical machine is violently deregistered (its VMs are terminated abruptly without migrating them to another machine), then it is expected that user-side schedulers will recover the lost work and make infrastructure faults transparent to their users.

\subsection{Output analysis techniques}

As the simulator is designed to be used with not-yet-known use cases, it was unknown what kind of output and state monitoring would be required for a particular use case. Therefore, the simulator offers state monitoring facilities throughout its components. But these monitoring facilities are not used by the simulator itself. This design decision allows the simulator to run with high performance when there are no direct observations needed by simulation developers. On the other hand, the simulation developers can create just about any kind of monitoring mechanism (mimicking their real life counterparts) on top of the offered facilities. Thus these custom monitoring mechanisms allow the collection of simulation outputs best tailored for the particular use case scenario of the simulation developer. 

\subsubsection{Output collection methods}

The simulator provides three techniques for monitoring the state of the simulated systems: $(i)$ direct one-time query, $(ii)$ periodic data collection with polling, and $(iii)$ notifications about important state changes. The first two techniques are expected to be utilised when the general state of the simulated system is required for decision making (e.g., setting up stop conditions for the developed simulations). In contrast, the last technique is more relevant for micro-management situations where immediate actions are needed in reaction for some temporal situations in the simulated system (e.g., creating triggers that increase the switched on physical machine count of a saturated infrastructure).

The following list provides an overview on some of the most relevant infrastructure state related one-time query facilities of DISSECT-CF:
\begin{itemize}
\item Total consumption recorded by a particular energy meter;
\item Total and available storage capacity of a cloud, repository or a physical machine;
\item Number of bytes sent/received by a network node;
\item Number of processed instructions by a physical/virtual machine;
\item Number of currently ongoing network transfers or processing tasks -- $\mathcal{P}(s,t)$ -- in a given resource spreader;
\item Total, freely available and currently allocated amount of computing resources in a cloud, physical machine;
\item Number of repositories, physical machines, virtual machines and queued VM requests in a particular cloud.
\end{itemize}
With the help of the periodic collection technique these metrics can reveal the progression of the simulated system's state. This periodic collection technique was one of the major design factors behind the recurring event mechanism (embodied in the \verb+Timed+ class) in the event subsystem of the simulator. This event mechanism allows the creation of simple but performant monitoring mechanisms over all direct one-time query facilities in DISSECT-CF. Basically, the simulator developer is expected to create a subclass of \verb+Timed+, implement the desired polling mechanism in its \verb+tick+ function and \verb+subscribe+ to recurring notifications at the event subsystem. Within this newly created polling mechanism the simulator developer can devise several statistical analysis tools and triggers (e.g., revise multi-layered scheduling strategies if the average number of allocated computing resources stay below a certain limit) as well as some aggregates for overall representation of the system state.

Complementing polling based mechanisms that mostly offer an overview on the simulated system's behaviour, DISSECT-CF also offers notifications for particular system states (i.e., for those states that would require alarmingly high polling frequencies). The following list provides an overview on the most relevant notifications offered by the simulator:
\begin{itemize}
\item State changes of an observed physical or virtual machine (the possible state transitions of virtual machines are shown in Figure~\ref{FIG-VMStates}).
\item For IaaS systems and physical machines, one can request notifications about the changes in the available processing capacities (e.g., the number of CPU cores). Additionally, for physical machines, one can even ask for notifications about released resource allocations -- i.e., terminating/migrating VMs of a physical machine.
\item Also, changes in the VM request queue can be kept under surveillance. Allowing simulation developers to adopt their user-side scheduling techniques to the state of the IaaS side VM queue.
\end{itemize}

\subsubsection{Evaluating the collected output}

The simulation developer will have a chance to use the collected outputs for two main purposes: defining the termination condition and analysing system behaviour. As the system behaviour analysis is expected to be done after the simulation was executed, these techniques are out of scope of this paper. The highly customisable periodic collection technique of the simulator allows storing the collected data in just the right format for the analysis software used by the simulation developer after data collection (the output and its format is expected to be defined by the developer). Therefore, the next paragraph solely focuses on the way the termination condition could be evaluated.

Simulations can be terminated with two approaches. With the first approach, the simulation developer is expected to ask the simulator to run the simulation until there are no further events in its queue -- this can be achieved by calling \verb+Timed.simulateUntilLastEvent()+. Thus, in this approach the event queue must be cleared by the simulation developer once the desired condition in the simulated system state is reached. The queue's cleanup will most likely be initiated by one of the periodic data collectors implemented by the simulator developer. In contrast to this approach, DISSECT-CF also allows the simulation developer to create complex termination condition analysis techniques which could be executed during the developer inserted gaps in the simulation. These gaps are created by simulating for a certain time period (by utilising the \verb+Timed.simulateUntil()+ function), then stopping for evaluation of the system state. The time period can be automatically adjusted based on the likelihood of reaching the simulation's completion condition within a given time frame. This approach allows less invasive state monitoring, or allows more complex termination condition evaluation techniques that would significantly reduce the simulation's performance if executed via periodic polling techniques.

\section{Evaluation} \label{sec-eval}

This section provides an analysis of the currently available version of DISSECT-CF. The section first starts with comparing small-scale real-life measurements to results received from simulations. During the comparisons, the accuracy of the simulation is analysed and several approaches are discussed that could increase the accuracy even further. In the later parts of the section, the article focuses on a comparative performance and accuracy study of DISSECT-CF with two existing simulators. During this study the most efficient simulation conditions for DISSECT-CF are shown through the analysis of its reactions to large-scale (over one million tasks) synthetic and real-life workloads. Finally, the section closes with the analysis of the performance impacts introduced by energy meters.

In the rest of the section, when results from any of the simulators are discussed, they were always obtained using a machine with the following components: $(i)$ an eight core Intel(R) Xeon(R) X5570 CPU   running at 2.93 GHz, $(ii)$ 32 GB of memory, $(iii)$ 128 GB SSD and $(iv)$ 10 Gb Infiniband network.

\subsection{Validation with real-life measurements}

\subsubsection{CPU sharing with parallel tasks}\label{SEC-CPUSHARE}

The first experiments were targeted at the CPU sharing mechanisms inside DISSECT-CF. For these experiments, two simple applications were developed with significantly different workload characteristics: $(i)$ CPU intensive and $(ii)$ memory intensive.

\paragraph{Behaviour of the CPU intensive workload} 
For the purpose of the evaluation, a simple workload function ($\mathcal{W}: \mathbb{N} \to \mathbb{R}$) was defined that can be evaluated iteratively:
\begin{equation} \label{EQ-Workload}
\mathcal{W}(i)=\left\{\begin{array}{ll}
i\cdot\mathcal{W}(i-1) & \textrm{if }i\textrm{ is even}\\
\mathcal{W}(i-1)/i & \textrm{otherwise}
\end{array}\right.
\end{equation}
Where the assumption of $\mathcal{W}(0)=1$ was made. As can be seen, to determine $\mathcal{W}(4)$ one first needs to evaluate the workload function with values 1-2-3. Thus the function scales linearly -- i.e., it takes twice as much computing power to evaluate $\mathcal{W}(2i)$ than $\mathcal{W}(i)$.

Using this linearity the CPU intensive experiment is executed as follows: $(i)$ as its input it receives the number of parallel threads to create ($TC$) and a starting $i_{min}$; $(ii)$ for each created thread it assigns a task number identifier ($TN_{id} \in \mathbb{N}$) ranging between zero and $TC$; $(iii)$ afterwards the threads will be asked to calculate $\mathcal{W}(i_{min}\cdot(TN_{id}+1))$; $(iv)$ all threads are started after all of them have received their tasks; and finally $(v)$ each thread prints out the time it takes to evaluate its assigned calculation. In the following experiments, the $i_{min}$ value was selected so a single-threaded workload function evaluation ($TC=1$) took around two-seconds to execute.

The experiment's construction allows the observation of three distinct scenarios: $(i)$ over-provisioning (if $TC-TN_{id}<cores(CPU)$), $(ii)$ resource utilisation balance (if $TC-TN_{id}=cores(CPU)$), and $(iii)$ under-provisioning (when $TC-TN_{id}>cores(CPU)$), where the $cores$ function evaluates the number of processing cores of the system ($CPU$) on which the experiments ran. Through these scenarios the simulator's default resource scheduling mechanism (see Section~\ref{Sec-URSM}) can be comprehensively evaluated.

\paragraph{The behaviour of the memory intensive workload} Instead of defining a new function, this workload only alters the way past-evaluated $\mathcal{W}(i)$ values are stored. While in the CPU intensive approach only the last calculated value is kept (in order to allow its storage in a register), the memory intensive workload stores all past values in the memory. Thus, when a memory intensive thread receives its task (see the 3$^{rd}$ step in the CPU intensive workload's description), it immediately allocates memory that could store all $i_{min}\cdot(TN_{id}+1)$ future values of the workload function (see Eq. \ref{EQ-Workload}). In the experiments, the applied $i_{min}$ value is also different from CPU intensive workloads: it was the maximum possible that still did not cause swapping on the system used for evaluation.

\paragraph{Experiments} For the CPU sharing experiments two machines were used with the following specifications: $(i)$ an AMD Opteron based, dual socket, eight-core machine with 32 GB memory and 512GB hard drive, and $(ii)$ a machine equipped with an Intel Core i7 processor (two cores running at 1.8GHz, hyper threading enabled), 4 GB memory and 256 GB SSD. The second machine was added to the experiments to show how the simulation can be adjusted to support hyper threading enabled processors. On both machines a single virtual machine was created and it occupied all CPU cores (in the case of the hyper threading enabled machine the VM was using four virtual CPUs), 75\% of the physical memory and 1 GB disk space. Both of the workloads were executed with a single thread first on the newly created virtual machines. With this execution, the baseline measurement was obtained for the single task length that was used in the simulator to set the remaining consumption ($p_r$) for each resource consumption sent to the simulated virtual machine.

\begin{figure}[tb]
\center
\includegraphics[width=\columnwidth]{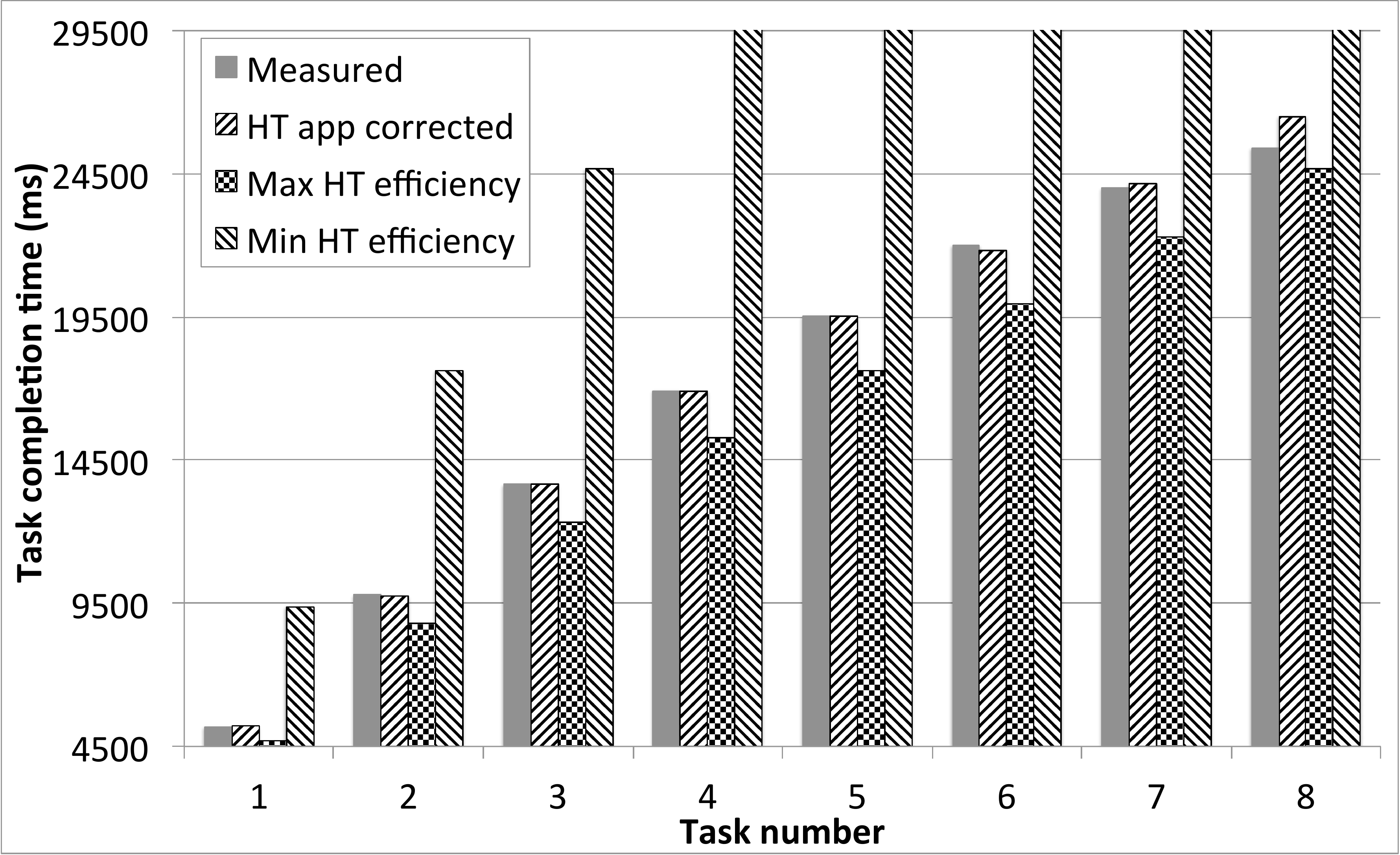}
\caption{Parallel execution of eight CPU intensive tasks\label{FIG-IncParTask}}
\end{figure}

In case of the \emph{CPU intensive workload}, the simulator was able to predict with a relative error of less than 0.1\% (with a standard deviation of 0.2\%) the completion time of each thread running on the virtual machine hosted in the Opteron based machine. On the other hand, as expected, the 4 virtual cores in the VM hosted by the Intel processor did not perform like 4 individual cores. Thus, a different set-up was needed for resource consumptions simulated on the non-independent cores of the Intel processor based virtual machine. Figure~\ref{FIG-IncParTask} shows three sub-experiments to analyse the effects of resource consumption set-up on the precision of the simulated results. In the first sub-experiment, the simulated VM was set-up as if the 4 cores would be just like the AMD based individual cores (the figure represents this simulation with ``\emph{Max HT efficiency}''). In the next sub-experiment, the 4 cores of the simulated VM performed just like a 2-individual-core based system (the figure represents this simulation with ``\emph{Min HT efficiency}''). Finally in the last sub-experiment, the processing limit $p_l$ for each defined resource consumption was customisable. Using this customisation facility, the processing limit was adjusted to a level that minimised the relative error of the simulated computing task runtimes. With the experimental workload, the applied limit was 89.6\% of the originally expected processing power of a single core. After applying such processing limit, the achieved task runtimes are shown with the label ``\emph{HT app corrected}'' in the figure.

The figure reveals that simulation developers of DISSECT-CF should not assume maximum or minimum hyper threading efficiency. Instead, if possible, they should evaluate their application workloads and when sending these workloads to the simulator, they should adjust the processing limit of the resource consumptions related to their application. As the figure shows, the adjusted workloads match the real-life measurements the most closely (i.e., the relative error is 0.29\%, and its standard deviation of is 1.49\%). Unfortunately, in cases of over-provisioning (see Tasks 5-8), the simulator's prediction performance decreases and in situations when significant over-provisioning is happening (e.g., with Task 8) the adjustment of the processing limit is not recommended (i.e., assuming 4 cores with maximum hyper threading efficiency produces better results). The accuracy has dropped because the less utilised a hyper threading based processor is, the more performance is achievable by a single thread. Despite this processor behaviour, the simulator still applies the previously set up processing limit in over-provisioning situations. To avoid this static behaviour in the current version of DISSECT-CF, simulation developers should dynamically change the processing capabilities of the PMs depending on the level of over-provisioning they observe.

\begin{figure}[tb]
\center
\includegraphics[width=\columnwidth]{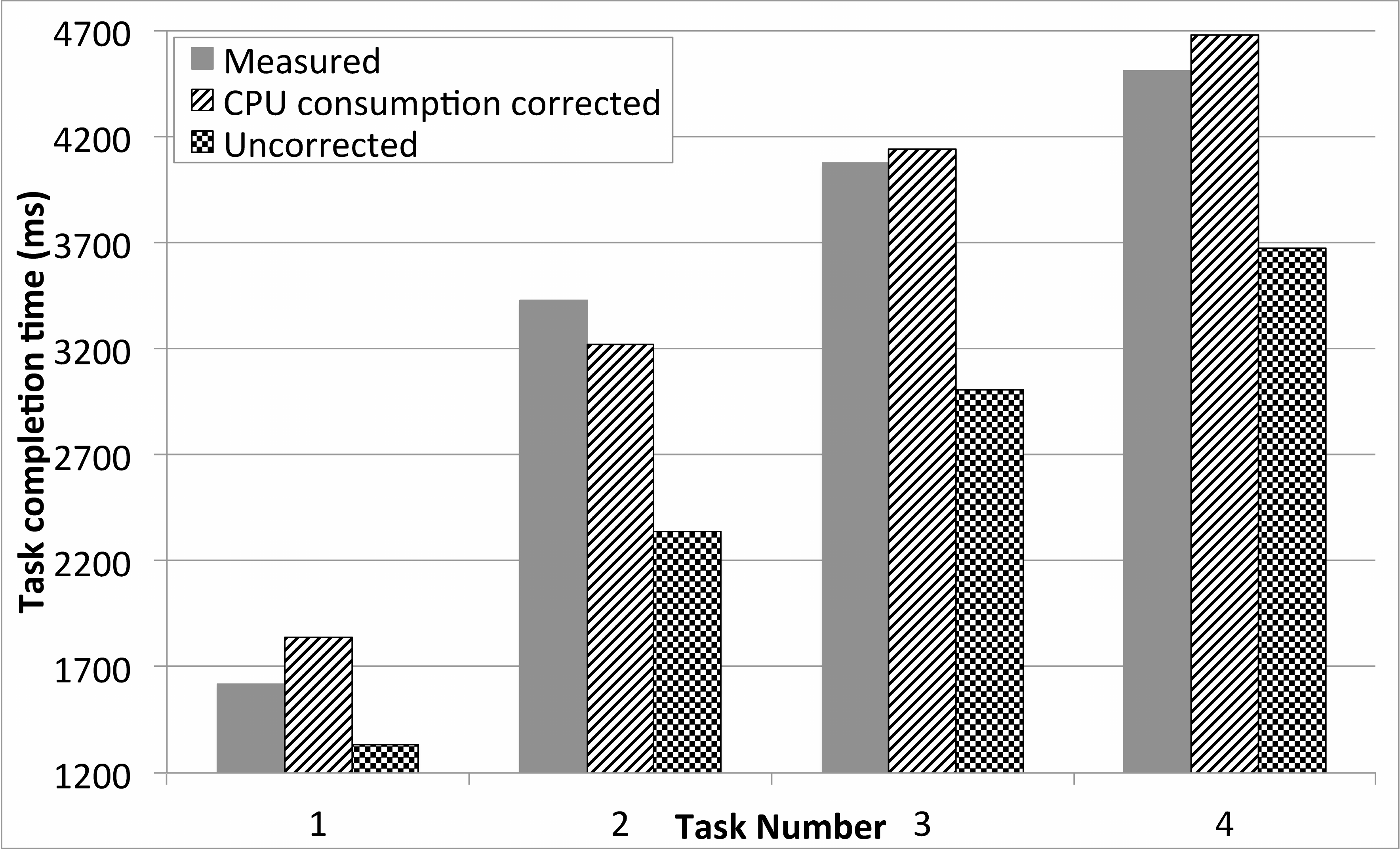}
\caption{Parallel execution of four memory intensive tasks\label{FIG-MemoryIntensive}}
\end{figure}

The case of the \emph{memory intensive workloads} is completely different. As DISSECT-CF in its current form does not simulate memory behaviour, the best one can do to get a close-to-real-life prediction from the simulator is to analyse the CPU load of the memory intensive application and adjust processing limits accordingly. Unfortunately, even slight changes to the memory access patterns of a workload could significantly alter prediction performance. Figure~\ref{FIG-MemoryIntensive} shows an example of how volatile the predictions could be. To present the volatility, first, a slightly different implementation of the memory intensive workload was created (this version did not allocate memory in advance but did the allocation as its memory needs increased). Then, the processing limits were optimised for this kind of workload similarly as it was done for the CPU intensive workloads. Next, the above-discussed memory intensive workload was executed. The runtimes of its four threads are represented with the ``\emph{Measured}'' columns. Afterwards, a simulation was executed with the previously identified processing limits (for the slightly different workload). The figure shows the simulator's runtime predictions for the four threads with the ``\emph{Uncorrected}'' column. These predictions were not useful (their relative error was over 22\%). On the other hand, if the processing limits were readjusted to match the memory intensive workload introduced a few paragraphs before, then the relative error can be brought within a more acceptable 4.75\%. These predictions of the simulator can be seen in the column labelled ``\emph{CPU consumption corrected}''.

\subsubsection{Networking}

Just like CPU sharing, network resource sharing is also based on the unified resource sharing mechanism of the simulator. But unlike CPU sharing, in network sharing situations there could be multiple resource providers that complicate the low-level resource scheduling mechanisms. In single provider scenarios, the findings of the previous subsection also apply to networking. Thus, this sub-section's focus is exclusively on the evaluation of a multi-provider situation when resource utilisation bottlenecks could occur because of the topology between the various providers and consumers.

\begin{figure}[tb]
\center
\includegraphics[width=0.63\columnwidth]{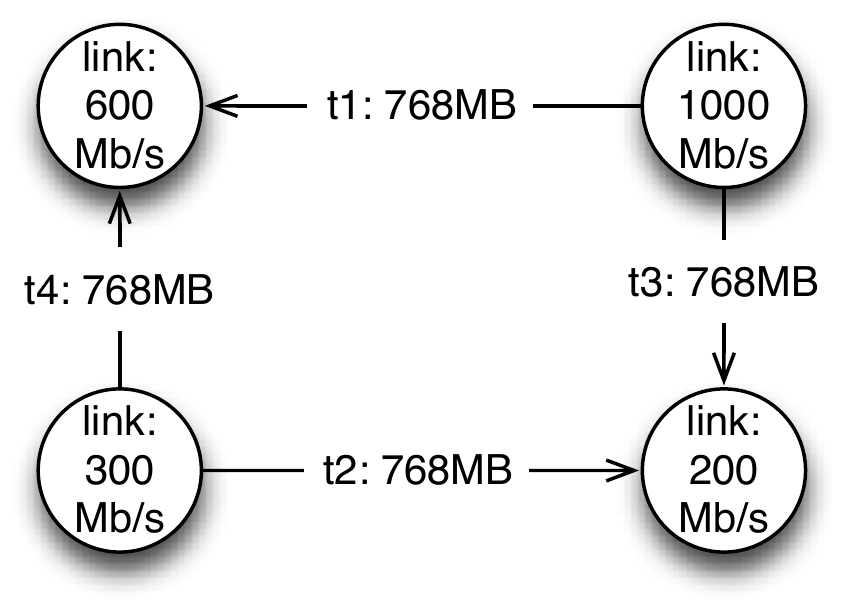}
\caption{Network interactions during the multi-provider evaluation scenario\label{FIG-NWTest}}
\end{figure}

Instead of a large-scale network setup, a simple and inexpensive network set up was chosen that can  experience bottleneck situations. The network was composed of 4 nodes with 1 Gb Ethernet connections to a managed gigabit switch. Their incoming and outgoing connections were shaped as shown in Figure~\ref{FIG-NWTest} (where the nodes are represented with circles and their maximum allowed incoming/outgoing bandwidth is shown inside the circle). Each node generated a 768 MB long completely random sequence of data that it stored in its local memory (this technique avoids any possible compression). The size of the generated data was selected to imitate a typical virtual machine image size in today's cloud infrastructures. The nodes then were instructed to transfer the generated data between them in a specific pattern. The requested transfers are depicted as arrows (starting from providers and aiming at consumers) in Figure~\ref{FIG-NWTest}. All transfers start at the same time and are accomplished using the HTTP protocol (widely used in the context of cloud storage).

The simulation was set up with two approaches: $(i)$ incoming and outgoing network connections were set up as resource spreaders with the maximum processing capabilities depicted in the figure; and $(ii)$ all network connections were set up as gigabit connections and the shaping mechanism was represented in the processing limits of the created resource consumptions (i.e., transfers) on the network. As expected, in both cases, the simulation results in the same transfer times for all transfers shown in the figure (\emph{t1} takes 15 seconds, \emph{t2} and \emph{t3} takes 60 seconds and \emph{t4} takes 30 seconds). Then the predicted timing results were compared to real-life measurements. The above-mentioned transfers were executed ten times and calculated the median completion time of each transfer in the real network. The relative error was within 0.5\% of the predicted timing results of the simulator.  

\subsubsection{Energy modelling}

As energy modelling depends on the resource sharing mechanisms of the simulation, its most accurate readings are bound to the accuracy of the sharing mechanisms. Previously, the most accurate predictions of the simulator were produced for CPU sharing on the Opteron based machine (see Section \ref{SEC-CPUSHARE}). Therefore, to present the best achievable accuracy for the linear power model of the simulator, the following measurement scenario was executed on the Opteron based machine.

\begin{figure}[tb]
\center
\includegraphics[width=\columnwidth]{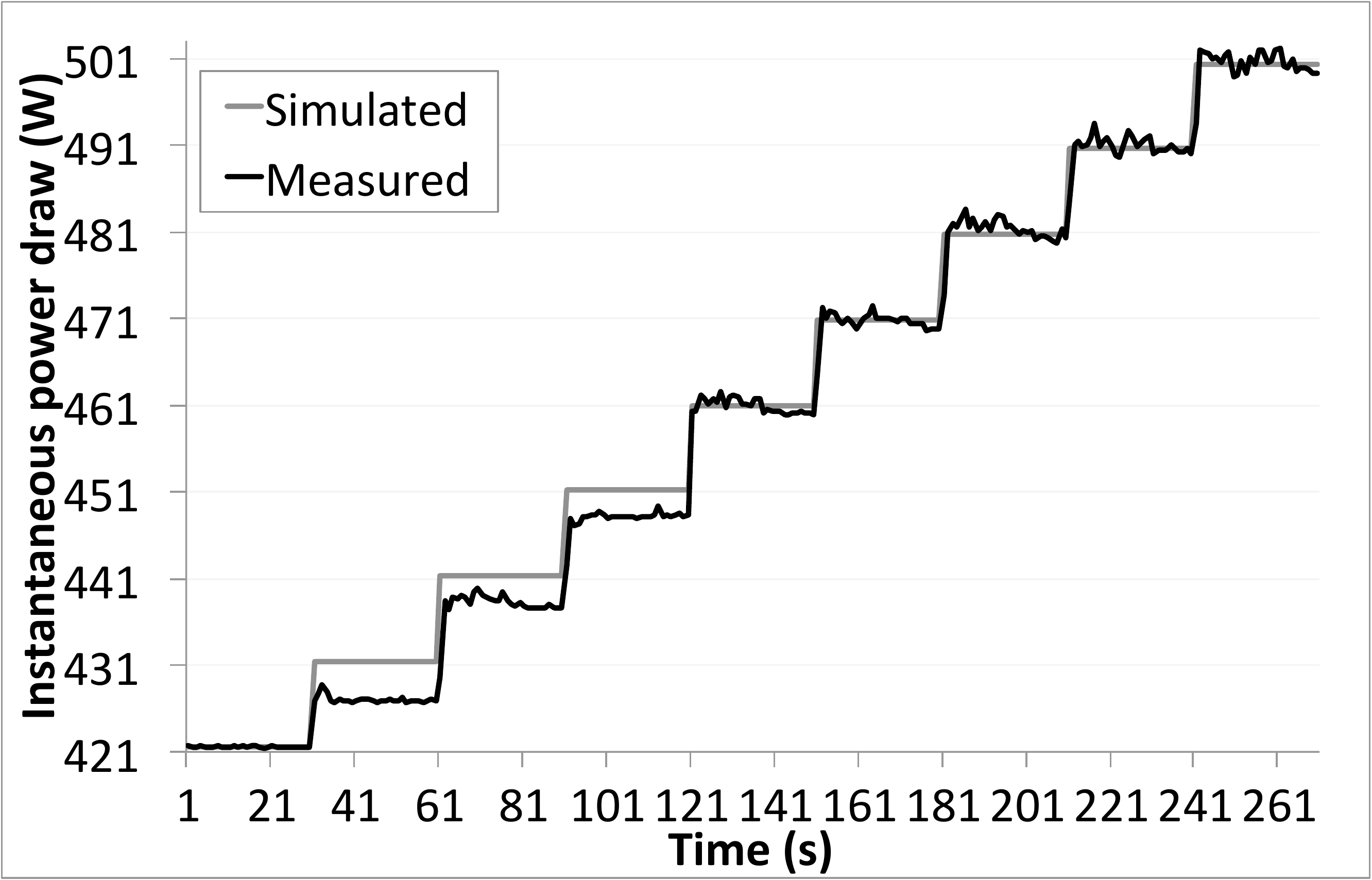}
\caption{Power estimation accuracy of the simulator\label{FIG-PWTest}}
\end{figure}

A new virtual machine image was created to host the application for the CPU intensive workload function (see Eq. \ref{EQ-Workload}). This image was transferred to the Opteron based machine and it was used to initiate eight virtual machines (all using a single core).  After the VMs started up and the machines reached their idle state, the power measurement was started. Next, in every 30 seconds one of the idle VMs started to evaluate the workload function for a sufficiently large $n$ value to utilise all the CPU resources of the selected VM for a period of over 10 minutes (i.e., this step ensured the overlapped utilisation of the physical machine's processor). The instantaneous power draw of the machine was collected during the entire experiment with a frequency of 2 Hz. The resulting power readings are presented in Figure~\ref{FIG-PWTest}. In parallel, the same scenario was also constructed in DISSECT-CF. The simulator was instructed to collect the instantaneous power draw values available from the linear \verb+Consumption Model+ (for details see Section~\ref{SEC-ENMET}). As the energy consumption model interpolates load between the idle and maximum energy consumptions, the best power draw prediction accuracy is expected in these two extreme cases. The collected power draw values are presented in the figure as a light grey line.

Confirming the initial assumptions, the relative error of the instantaneous power draw values are higher than in the case of CPU resource sharing. The observed relative error was 0.21\% with a sample standard deviation of 0.4\%. This means that the relative error was increased by a factor of over two. Consequently, the relative error in power readings could be significant in cases when the resource sharing mechanism is already experiencing larger errors. To avoid the loss of precision in such cases, it is expected that the factor of the relative error can be further reduced with energy consumption models that utilise better power interpolation techniques or exploit additional information about the system setup -- e.g., the number of CPU sockets.

\subsection{Performance comparison with other simulators}

In order to compare the performance of DISSECT-CF to its competition, several currently available simulators were investigated for similarity to the external interfaces of DISSECT-CF (see Section~\ref{SEC-UI}). Because of the similar interfaces the development time of performance evaluation can be reduced significantly. Fortunately, these interfaces not only reduce development time, but also avoid the need to implement similarly behaving components to other simulators. Several simulators were ruled out primarily because these additional components would significantly add to their functionality and therefore ruin their comparability with DISSECT-CF. According to these criteria, the following two simulators were selected: $(i)$ CloudSim 3.0.3 -- which is amongst the most widely used simulators for cloud systems~\cite{CloudSim-buyya2009modeling} --, and $(ii)$ GroudSim 2.0 -- a high performance simulator for grids and clouds developed at the University of Innsbruck~\cite{GroudSim-ostermann2011groudsim}. Furthermore, these simulators were chosen because of their accessibility at the time of the performance evaluations.

\paragraph{The prepared simulation environments} All simulations were run on the machine mentioned in the introductory paragraph of this section. As Java garbage collection could introduce performance differences because of the various simulators' memory usage patterns, the physical machine's entire memory (32GB) was allocated to the Java Virtual Machine (JVM). With this technique the garbage collector was practically never run and therefore allowed disruption-less evaluation of the various simulators.

Because of the way DISSECT-CF is constructed, it completely avoids any kind of logging mechanisms. If a simulation developer needs some logging mechanism, DISSECT-CF expects that he/she will attach to the necessary event handlers of the simulator and focus logging on his/her purposes. Unfortunately, the two other chosen simulators do not apply this technique. To ensure a fair comparison, before running any of the simulations all of their logging mechanisms were disabled (in some rare cases this needed non-functional changes to the GroudSim's code).

Several techniques (like internal investigation of the various simulators and additional event handling techniques) were applied to reduce the effects of the developed performance evaluation code on the various simulations. Before taking measurements, all simulations were evaluated with Java's embedded CPU sampler. The CPU share of the performance evaluation code varied from 0.5\% to 1.3\% in the worst case. In practice, this means that all evaluations and measurements reported in this section spent 99\% of their time inside the particular simulator (thus measurement differences can be rooted in the internal behaviour of the variuos simulators mainly).

\subsubsection{Evaluation with synthetic loads} \label{SEC-SYNTHEVAL}

Synthetic loads were used to evaluate the unified resource sharing model of DISSECT-CF. As the selected simulators mostly focus on CPU sharing, the following evaluations were focused on CPU sharing as well. In addition, DISSECT-CF applies the same unified sharing model to network and disk resources. Thus, the performance evaluations presented below are expected to be applicable to such resources, too.

\paragraph{Evaluation setup} In all simulators, a single core virtual machine was started up and ready to accept tasks. In GroudSim and CloudSim this requires the definition of a cloud/data centre and requesting a VM from it. In DISSECT-CF this step requires the definition of a physical machine booting it, placing a runnable virtual machine image into its local disk and starting that image as a VM.

\begin{figure}[tb]
\center
\includegraphics[width=0.8\columnwidth]{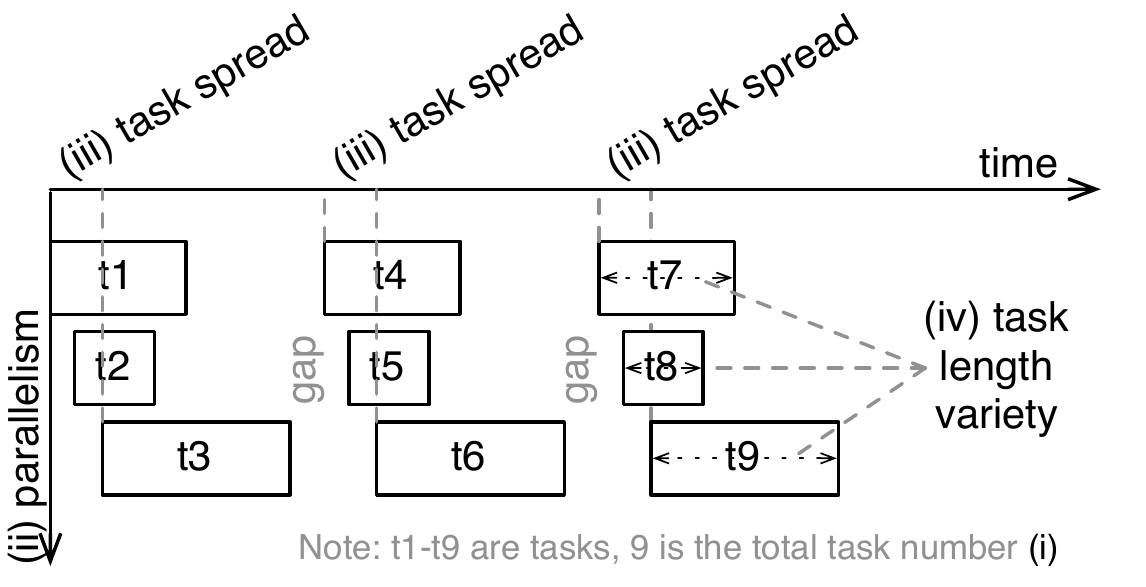}
\caption{Characterisation possibilities of the synthetic load generator\label{FIG-SLChar}}
\end{figure}

After the VM was set up, an artificial trace was generated with the following properties: $(i)$ number of tasks to be executed, $(ii)$ number of parallel tasks possible at once, $(iii)$ spread of the parallel tasks (within how much time all parallel tasks must start) and $(iv)$ the length variety of each possible task (randomly picked from a range). These properties are visualised in Figure~\ref{FIG-SLChar}. The number of tasks was selected for each simulation to be high enough that a single simulation run took more than 10 seconds. This technique filters out the temporal effects of Java start-up on the performance of the simulation. If the number of tasks is more than the number of parallel tasks possible, then the trace generator will insert a gap long enough for all the previously generated tasks to finish, and add a new set of parallel tasks to the trace. The trace generator ensures that each task added will exhaust all the CPU resources of the virtual machine for its entire duration. The rest of the trace properties will be discussed in detail during the evaluation.

Finally, after the trace is generated, the measurement starts by executing the trace on the virtual machine. The measurement stops when the complete generated trace has run. Each measurement records the time passed in nanoseconds between the submission of the trace's first task and the completion of the trace's last task.

\paragraph{Resource sharing performance}

\begin{table*}[tb]
\centering
\caption{Number of tasks to run for at least 10s long simulation time\label{TAB-TASKDISTRS}}
\begin{tabular}{c|rrrrrr}
\hline
\multirow{2}{*}{Simulator}& \multicolumn{6}{c}{Parallel task number}\\
& 1 & 10 & 100 & 1,000 & 10,000 & 100,000\\
\hline
DISSECT-CF & 15,000,000 & 6,000,000 & 5,000,000 & 1,000,000 & 200,000 & 100,000\\
GroudSim & 5,000,000 & 2,000,000 & 200,000 & 25,000 & N/A & N/A\\
CloudSim & 5,000,000 & 5,000,000 & 200,000 & 1,000 & N/A & N/A\\
\hline
\end{tabular}
\end{table*}

\begin{figure}[tb]
\center
\includegraphics[width=\columnwidth, trim=2.4cm 10.7cm 7.8cm 3.2cm, clip=true]{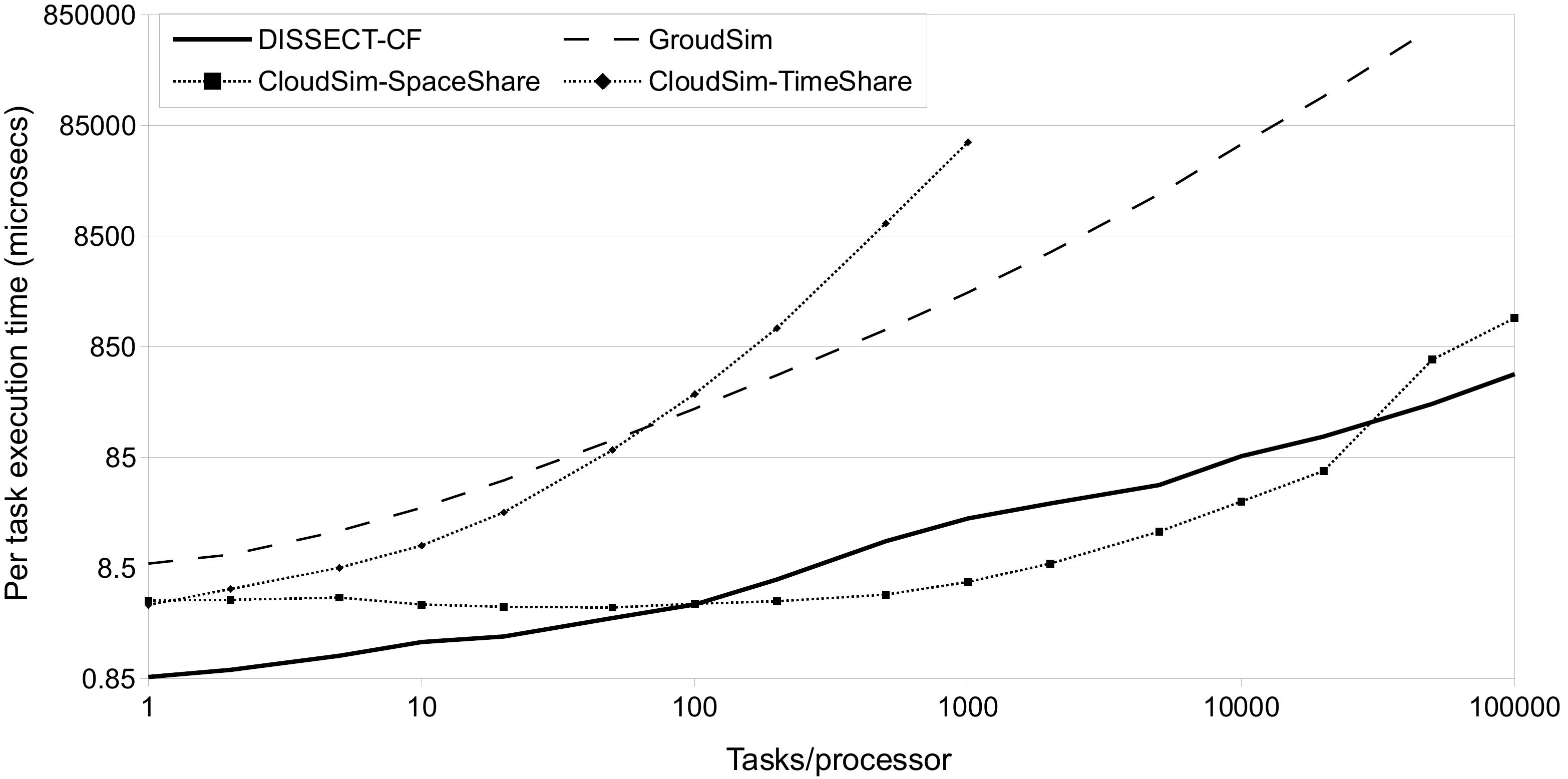}
\caption{Pure resource sharing performance in the studied simulators\label{FIG-PJRuntime}}
\end{figure}

During the performance evaluation, all parallel tasks of the trace started up in the first 10 seconds (\emph{task spread}) and had a length variety between 10-90 seconds. The necessary amount of tasks for the 10-second-long simulation runtime is shown in Table~\ref{TAB-TASKDISTRS}. Each evaluation was ran with different parallel task numbers between (1-100,000) to allow the investigation on how the increasing parallelism changed the performance of the simulators. The findings are shown in Figure~\ref{FIG-PJRuntime}.

As the figure shows, in the case of CloudSim, both its time-shared and space-shared VM scheduler was experimented on. The time-shared scheduler of CloudSim simulates parallelism, while the space-shared scheduler serialises the arrived tasks so at any given time there is only one task that can utilise the CPU, and the rest are queued until this task is finished. Unfortunately, the time-shared scheduling mechanism of CloudSim is broken; it lengthens task execution times significantly even in the case of a two-parallel-task setup (e.g., a simple setup like the one presented in Figure~\ref{FIG-IncParTask} leads to completely different results than one would see in real-life) On the other hand, the space-shared scheduler provides completely different task completion times (because of its serialising behaviour). Thus the CloudSim related details of the figure are only valid when considering their simulation runtime; the simulated tasks do not finish at the expected times in either case.

As depicted in the figure, CloudSim-time-shared and GroudSim measurements abruptly finish at 1,000 and 50,000 tasks. For higher levels of parallelism, the runtime of the simulation took more than 8 hours and therefore was cancelled. In the case of CloudSim, the performance penalties mainly originate from its centralised design of data centres (i.e., most of the logic and event processing reside in the \verb+Datacenter+ class, and the rest of CloudSim's classes are used for state representation only). GroudSim, on the other hand, has a different design issue: it pre calculates all task completion times, puts them into the event queue and thus if a change is needed to them, the whole event queue has to be updated.

Finally, for all simulators, the task processing performance shown in the figure is expected to be faster than what one usually would see, because of the large amount of tasks that were executed for gathering data for the figure. The amount of repeated evaluations allowed the JVM to optimise the runtime behaviour of each simulator for the particular performance evaluation scenario. Later on this effect is excluded from the evaluations.

\paragraph{Performance influence of load characteristics}

The rest of this sub-section analyses of how the performance of the resource sharing varies depending on changes in task arrival and length characteristics. For this analysis, a baseline measurement was collected with no parallel tasks but varying task lengths. Than the assumption was made that from this baseline, the simulators would linearly degrade in performance (i.e., two times the parallelism would increase the simulation run time of a single task by two times). To compare how linearly a particular simulator behaves under particular load characteristics, the following scaling ratio function was designed:
\begin{equation} \label{EQ-SCRATIO}
\mathfrak{s}(k,\rho,\mathfrak{n},d)=\frac{\sigma(k,\rho,\mathfrak{n},d)}{\mathfrak{n}\cdot\sigma(k,\rho,1,d)}
\end{equation}
Where $\mathfrak{s}$ is the scaling ratio, $k$ is the kind of simulator, $\rho$ is the chosen range of task length variety (in the below detailed experiments it was either 10-90 s or 200-3600 s), $\mathfrak{n}$ is the number of parallel tasks, $d$ is the task spread (in particular, 10 s or 200 s were used), and $\sigma$ is the measurement function that evaluates the particular simulator with the given load characteristics.

\begin{figure}[tb]
\center
\includegraphics[width=\columnwidth, trim=2cm 8.3cm 7.7cm 3.1cm, clip=true]{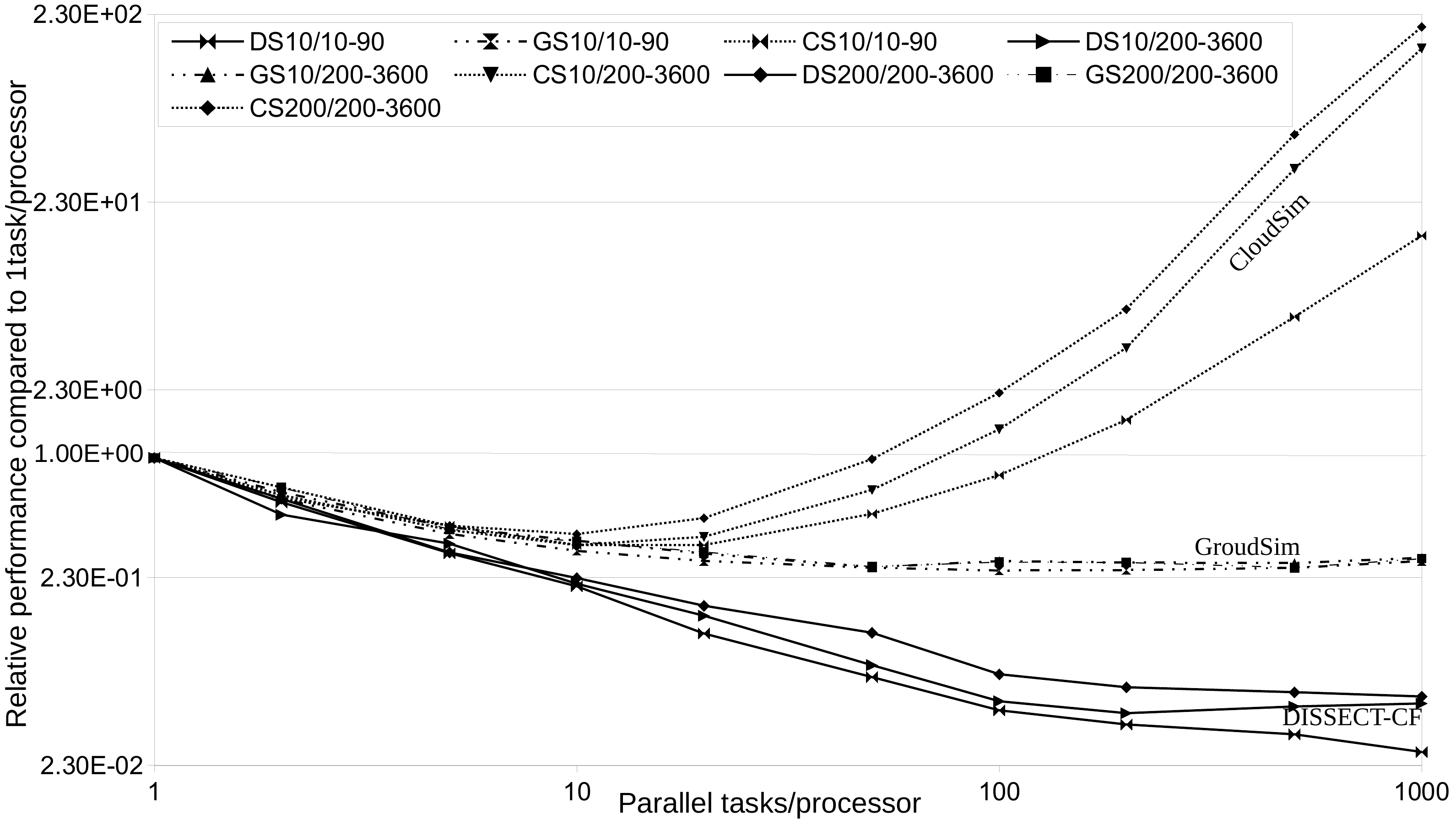}
\caption{The dependence of relative resource sharing performance on load characteristics in the investigated cloud simulators\label{FIG-JobScalingCompare}}
\end{figure}

Figure~\ref{FIG-JobScalingCompare} compares DISSECT-CF with the two selected simulators via the scaling ratio function. The labels of the figure are presented in the following format: $k\textrm{ }d/\rho$ (see Eq. \ref{EQ-SCRATIO}), where $k$ is one of DS/GS/CS, which translate to DISSECT-CF/GroudSim/CloudSim, respectively. The x-axis of the figure shows the increase in $\mathfrak{n}$.

Based on the distribution of the measured scaling values for the different load characteristics, the most tolerant to the change in workload is GroudSim, while the least tolerant is CloudSim. Degrading performance in scaling can be also observed for both CloudSim and GroudSim. In the case of CloudSim, the degradation starts around 10 parallel tasks, and by the time the simulation reaches 50 parallel tasks the simulator becomes worse than linear ($\mathfrak{s}>1$). In the case of GroudSim, the degradation starts around 100 parallel tasks, and becomes worse than linear around 20,000 parallel tasks. In both cases the early degradation is caused by the data structures and the indirect call structures used (i.e. often these simulators experience rather deep call stacks during parallel event handling). Finally, although it is not visible in this chart, the degradation starts for DISSECT-CF at around 10,000 parallel tasks, and based on estimates, its scaling becomes worse than linear around 1.5 million parallel tasks. Fortunately, this huge amount of parallelism is unlikely for most simulations. But simulations of some highly under-provisioned systems might need levels of parallelism that could lead to degraded performance even in GroudSim.

The figure also reveals that the performance of DISSECT-CF is more dependent on the task spread, while CloudSim's scaling is limited more by the task length variety. In case of DISSECT-CF, the task spread dependency manifests because the simulator executes its resource sharing mechanism only once per \SMALLESTIMEGRANULARITY, on new resource consumption arrivals and departures (see Figure~\ref{FIG-RS} for details). However, the increased task spread decreases the likelihood that the resource sharing mechanism can be executed on multiple resource consumptions at once. Therefore, the wider the spread the closer one gets to the worst possible resource sharing performance in DISSECT-CF. In fact, the 200 s long spread was used because this spread is already high enough to reveal close-to-worst-case performance. Increasing the spread further did not introduce significant resource sharing performance drops, but its impact on CloudSim based simulations rendered the evaluation of the 1,000 parallel task experiment too time consuming.

\subsubsection{Evaluation with loads similar to real-life}

As synthetic loads often criticised because of their possible bias, DISSECT-CF was evaluated and compared using workloads that were collected from real-life computing infrastructures. Thus, the evaluation required workload traces with the following characteristics: $(i)$ collected for extensive periods of time (i.e., at least a few months long) to ensure the widest variety of observable load characteristics for the particular infrastructure; $(ii)$ tasks should be described in detail including their resource utilisation and submission, start and completion times (so even if just task definitions are available, one can still translate them to the kinds of virtual machines the tasks would need for their execution in a cloud environment); and $(iii)$ if the traces contain virtual machine management logs, then task allocation details are also necessary to enable the analysis of new scheduling techniques that might aim at reallocating tasks or that would change VM management operations.

Based on these requirements, most of the traces (e.g., PlanetLab) containing virtual machine management logs were not found suitable for the planned comparative study; the rest of these log based traces are not collected for enough time to be used in large scale experiments. Thus, although these virtual machine management log based traces would be the best candidates for analysing cloud characteristics, their immaturity necessitates to also search amongst traces collected from other large-scale infrastructures like grids. Two appropriate sources were identified: the Grid Workloads Archive (GWA \cite{Iosup08thegrid}) and the Parallel Workloads Archive\footnote{http://www.cs.huji.ac.il/labs/parallel/workload/swf.html}. Because both at the University of Innsbruck and in MTA SZTAKI, there are earlier good experiences with the processing of GWA traces, this article presents a comparative study of the three selected simulators through the traces downloadable from GWA (namely: DAS2, Grid5000, NorduGrid, AuverGrid, SHARCNet, LCG). 

\paragraph{Trace processing} A trace loader was prepared for all three simulators so events were fired every time a task arrives. Every time a new task arrival event is fired, the simulators are programmed to create a virtual machine that will host the task. Once the VM was created the task was instantiated in it according to its definition in the trace file. When the task was completed according to the particular simulator, its hosting VM was also terminated. Because of the known problems with the time-sharing mechanism in CloudSim, single VMs did not receive parallel tasks (e.g., by requesting a VM that is sized to host multiple tasks first). Instead, when parallel tasks were needed, multiple VMs were created in the simulated cloud infrastructure.

Unfortunately, because of a few conceptual differences between DISSECT-CF and the other analysed simulators, the above-mentioned trace processing technique has several minor differences in their adaptations to the various simulators. First of all, DISSECT-CF has a queuing first-fit VM scheduler (see Section~\ref{Sec-HLS}) that allows users to send VM requests right upon task arrival and wait for the VM scheduler's queuing mechanism to notify them about the VM's running state (and readiness to receive the task). Unfortunately, the rest of the simulators do not offer VM request queuing: they reject VM requests that cannot be hosted according to the actual state of the simulated infrastructure. To have a similar behaviour to DISSECT-CF, either the VM request needs to be resent until it can be fulfilled (which would introduce an unwanted busy waiting loop), or alternatively one must rely on user-side information. If the VM request is repeatedly resent, then the other simulators are significantly disadvantaged. Thus, the second approach was used: previously non-servable VMs were only re-requested once one of the previous VMs have terminated (i.e., it was assumed that one can determine the state of all the running VMs at a given time instance). With this technique the CPU share of the performance evaluation code (as measured by Java's embedded CPU sampler) was kept around the same levels as it was for DISSECT-CF.

Tasks, which are utilising several CPU cores in parallel, are modelled in two ways. First, the simulation tries to request a VM with as many CPU cores as the task needs. Unfortunately, this cannot be achieved in cases when even the biggest possible VM is too small for the task's needs. For these huge tasks, the simulation requests multiple VMs, enough to fulfil the parallelism in the task. In GroudSim and DISSECT-CF it is possible to request multiple VM instances at once (e.g., similarly to how Amazon EC2 behaves) and they will be ensured to be available in parallel. On the other hand, CloudSim does not have such functionality; therefore it was extended with a technique that only submits the tasks to their virtual machines once all necessary VMs are available for the level of parallelism needed for the task.

As GroudSim is more focused on the user-side behaviour of cloud infrastructures, it has several deficiencies compared to the other two evaluated simulators: $(i)$ it can only handle tasks that occupy a single CPU core, $(ii)$ its network sharing mechanism could result in network under utilisation if the communicating parties are using connections with different bandwidths and $(iii)$  it does not provide data centre level simulation details -- e.g., no VM/PM scheduling and PM level resource sharing is simulated. To overcome the first deficiency, multi-core tasks are simulated as several single core tasks in the same GroudSim virtual machine. To avoid the problem with network sharing, the simulated data centres in all three simulators were constructed on such a way that every node was connected with the same bandwidth to the others. Unfortunately, without significantly changing GroudSim's code it was not possible to add the missing data centre level simulation details. Thus, during the performance analysis one should keep in mind that these details are not simulated in GroudSim.

Finally, for both GroudSim and CloudSim, the instantiation time of a virtual machine is instantaneous, which is not realistic. As the transfer of the VM image is often the most time consuming operation in the VM instantiation procedure~\cite{DISS}, this transfer was simulated with a download operation to the VM with both of the simulators. In GroudSim, a network transfer was initiated right after the VM was created and this transfer delayed the creation of the task on the VM until the transfer's  completion. In CloudSim, Cloudlets (computing tasks in CloudSim terminology) could have input files defined for them. Thus, tasks in CloudSim were specified so they must transfer an input file with the size of the VM image before they can start their processing. Unfortunately, even with input files specified, the VMs in CloudSim start immediately and execute their tasks right after the VM is created. This means that CloudSim based simulations cannot be as accurate as the other two simulator's results. In order to reduce the impact of VM image transfer on task execution times -- and to allow all three simulations to have a similar simulation completion time --, the size of the VM image was set to 100MB (which is the size of a rather small image nowadays).

\paragraph{The simulated virtual infrastructure}

In all three simulators, a single kind of physical machine acted as the foundation of the simulated infrastructure. This physical machine was modelled after a single node in MTA SZTAKI's cloud\footnote{http://cloud.sztaki.hu/en/home} and had the following properties: $(i)$ 64 CPU cores, $(ii)$ 256 GB RAM, $(iii)$ 5 TB local disk, and $(iv)$ two 1 Gb/s Ethernet connection. In the experiments detailed below, it was possible to define how many of these machines should be in the data centre. The set up of the machines includes the creation of a network interconnecting all of them with a central switch.

\paragraph{The runtime comparison experiment}

For the first experiment the simulated infrastructure was set up so it was sufficient to host even the largest parallelism a single task of the traces could request (i.e., no task has had to be dropped because there were not enough physical machines to host its level of parallelism). In particular, this experiment required the simulation of 20 physical machines with the above-mentioned properties. On this infrastructure, the first $n\in\mathbb{N}$ tasks from a particular trace were submitted. Then, a measurement was initiated for the real time passed between the submission of the first task and the completion of the last one. The infrastructure preparation, task submission and time measurement operations were repeated 10 times, each time starting with a new JVM. Then, the trace specific average simulation duration for the measurement of $n$ tasks was calculated. Afterwards, the whole measurement procedure was repeated for the remaining traces. Finally, the \emph{aggregated simulation duration} was calculated for $n$ tasks by averaging the trace specific averages from all traces. Later, the aggregated simulation duration is referred as $\mathcal{M}(k,\mathfrak{n},mc)$ (where $k$ is the measured simulator, $\mathfrak{n}$ is the number of tasks, $mc$ is the number of physical machines to simulate and $M: K\times \mathbb{N}^2\to\mathbb{R}$).

\begin{figure}[tb]
\center
\includegraphics[width=\columnwidth, trim=2.5cm 5.2cm 6cm 3.3cm, clip=true]{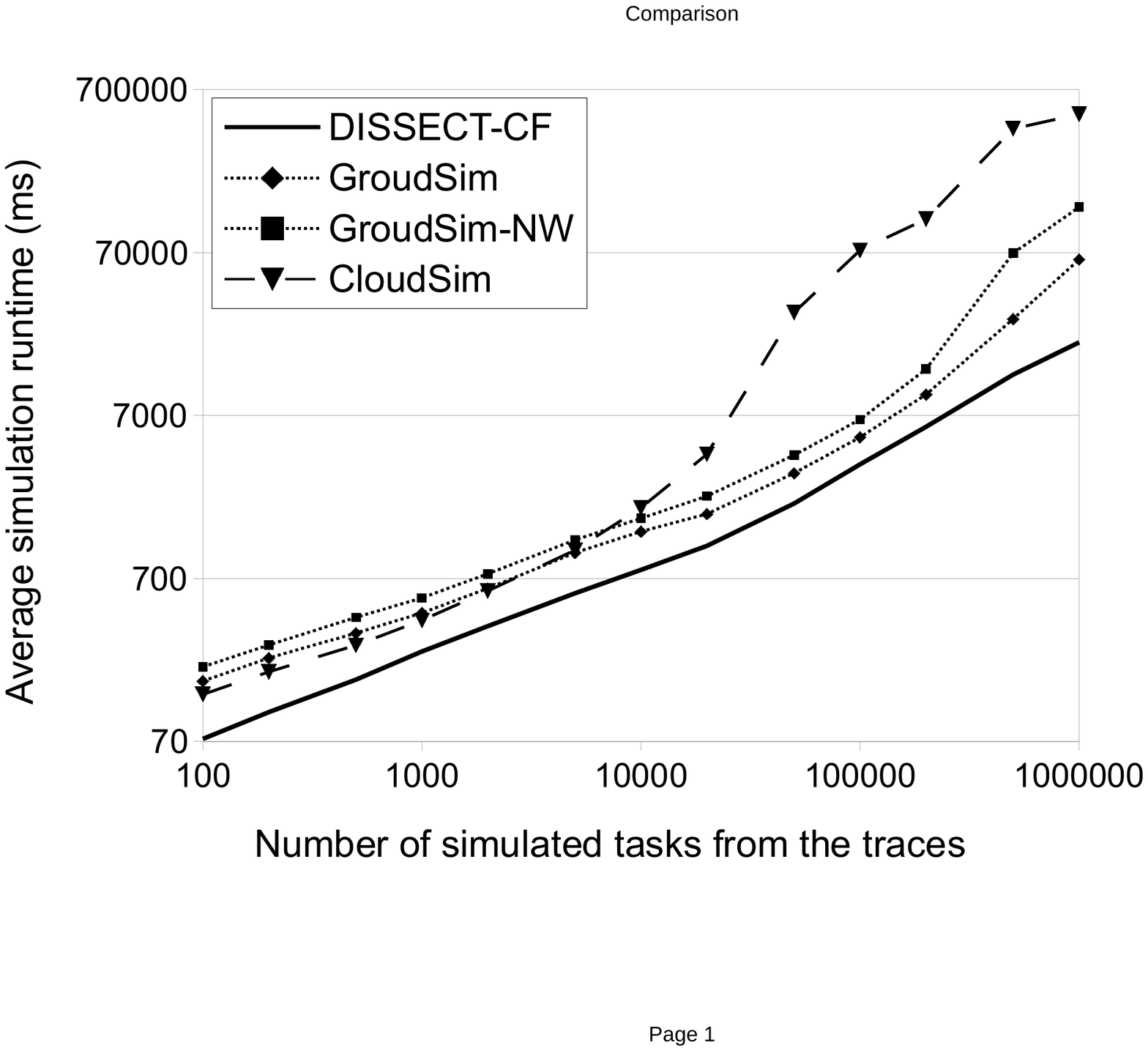}
\caption{Comparison of average runtimes of grid workload archive traces based simulations\label{FIG-RuntimeCompare}}
\end{figure}
\begin{figure*}[tb]
\center
\includegraphics[width=\textwidth, trim=2.3cm 7.7cm 4.5cm 3.2cm, clip=true]{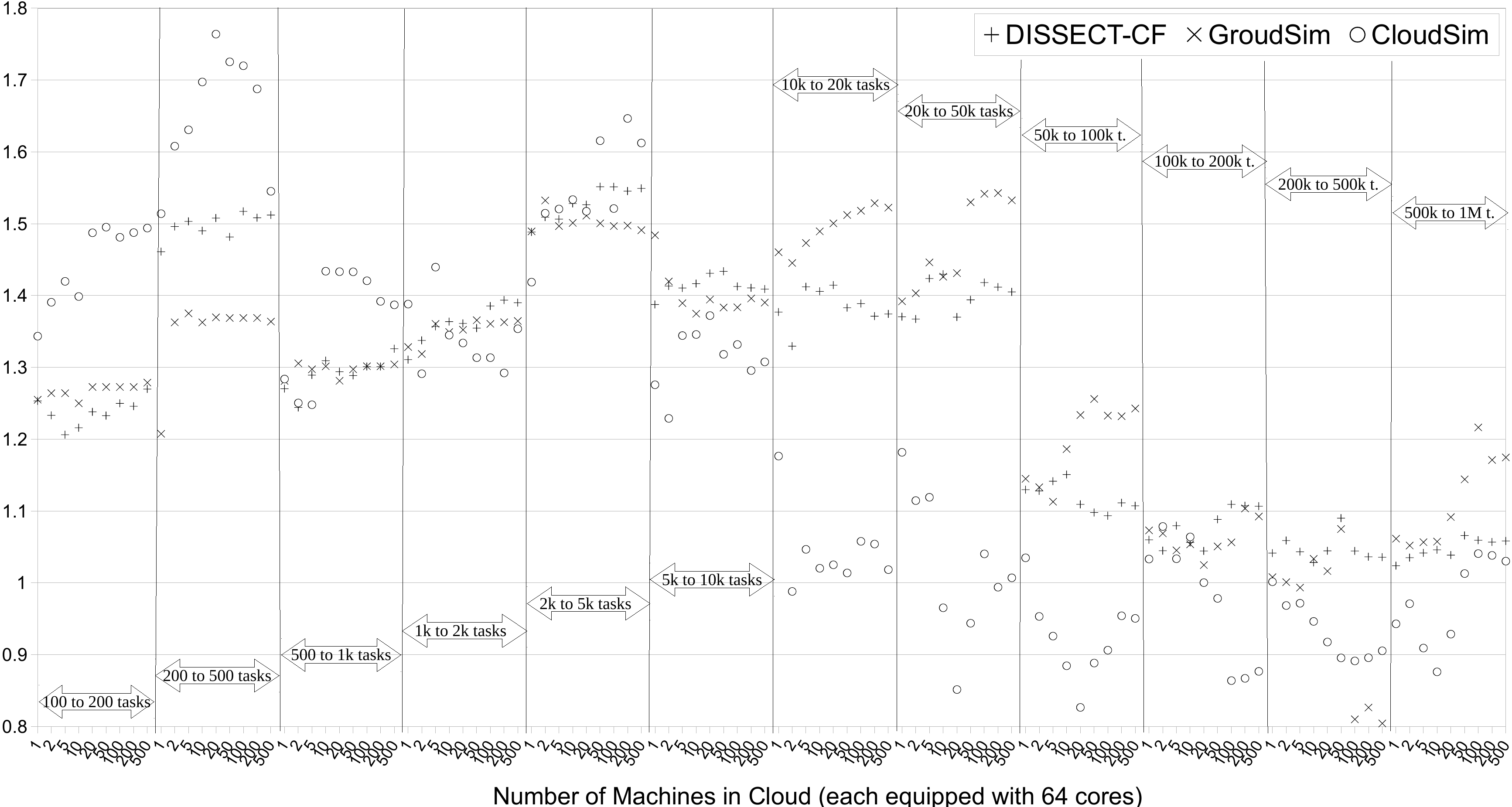}
\caption{Scaling comparison of DISSECT-CF \label{FIG-ScalingCompare}}
\end{figure*}

Figure~\ref{FIG-RuntimeCompare} presents these aggregated simulation durations while the number of tasks -- $\mathfrak{n}$ -- were changed from a hundred to a million. As the measurements ran with a cold JVM, for the first 10,000 tasks one can see that the JVM's effects on the start-up dominate the runtimes. The first workload related differences of the simulators can be seen after the JVM's behaviour is no longer a significant contributor to the measurement. Also, despite DISSECT-CF's focus on the internals of the infrastructure, its performance is always better than that of the other two simulators. This is especially apparent after around 200,000 tasks. In spite of GroudSim doing the lightest-weight simulation (see the deficiencies listed in the previous paragraphs), it is still significantly and consistently slower (albeit in some cases -- like the LCG trace -- GroudSim performs consistently better). The figure also shows that GroudSim's network simulation code gives around 21\% overhead for lower task counts, and for higher task counts the overhead could grow as high as 109\%. This difference experienced during the network simulation is caused by GroudSim's sub-par network modelling (i.e., for a single transfer to occur there are several events that needs to be handled in the event system, and in case of parallel transfers these events must be refreshed in the future event queue).

\paragraph{The scalability experiment}

To determine the scalability of the compared simulators, the previous experiment was repeated with varying sizes of infrastructure under them. Thus, these experiments will not only be able to pinpoint how the various simulators deal with the increasing amounts of tasks and virtual machines, but also how the simulators cope with increasing size of infrastructures. Infrastructures as small as a single physical machine were used but the experiments were reaching to the size of 500 machines. As even smaller infrastructures were evaluated than 20 machines (which was the minimum for the level of parallelism found in some of the traces), there were tasks that could not fit to the infrastructures. These tasks were automatically filtered out by the trace processor and never reached the simulation. Interestingly, even with the simulation with a single machine the filtered out tasks were never more than 6\% of the total number of tasks in any of the traces. Of course, the single machine still caused a significant serialising effect that lengthened the total execution time of the rest of the tasks.

Similarly to Eq. \ref{EQ-SCRATIO}, the scalability of the simulators was calculated compared to linear scaling. Figure~\ref{FIG-ScalingCompare} presents how the various simulators scale under a particular number of physical machines. The figure is composed of twelve sub-charts, each delimited with a black line. Between the black lines a large bidirectional arrow contains the title of a sub-chart. The title shows the number of tasks for which the final runtime measurement values were compared. For example: the title ``\emph{100-200 tasks}'' means the sub-chart shows how the final runtime measurement value of 200 tasks compares to the value of 100 tasks. The comparison is made according to the following equation:
\begin{equation}
\mathfrak{s}(k,\mathfrak{n_1},\mathfrak{n_2},mc)=\frac{\mathfrak{n_2}\cdot\mathcal{M}(k,\mathfrak{n_1},mc)}{\mathfrak{n_1}\cdot\mathcal{M}(k,\mathfrak{n_2},mc)}
\end{equation}
Where $\mathfrak{n_1}$, $\mathfrak{n_2}$ is the number of tasks for which the aggregated simulation duration is evaluated, $k$ is one of the evaluated simulators and $mc$ is the physical machine count. Thus, $\mathfrak{s}(k,\mathfrak{n_1},\mathfrak{n_2},mc)$ will become one if the particular simulator scales linearly in the task number range of $\mathfrak{n_1}$ and $\mathfrak{n_2}$. If the scaling function above is lower than one then the particular simulator is worse than linear. 

According to the sub-charts in Figure~\ref{FIG-ScalingCompare} all simulators scale significantly better than linear with fewer than 10,000 tasks (again, this is mostly due to the JVM's class loading and start-up mechanism, since practically all measurements under this task count result in sub-second runtimes). Afterwards, one can see that CloudSim is already scaling significantly worse than the other two investigated simulators, and by 50,000 tasks CloudSim already drops below linear scaling. It can be also observed that simulators handle the number of parallel machines completely differently. In the case of CloudSim, the bigger the machine number, the more likely that the scaling factor will become lower. In the case of GroudSim, the opposite behaviour is observed; while for DISSECT-CF, one can see a rather balanced case. In conclusion, GroudSim behaves as expected: the missing simulation of physical machine behaviour and VM/PM scheduling techniques allows it to behave practically independently from infrastructure size. In this sense, CloudSim and DISSECT-CF should behave more closely to each other. Unfortunately, the problematic implementation of CloudSim's time-shared VM scheduler (see Section~\ref{SEC-SYNTHEVAL}) changes its apparent behaviour and reduces the similarities in the trends between DISSECT-CF and CloudSim. To conclude, DISSECT-CF scales comparably to other state-of-the-art simulators (in fact it never drops below linear scaling, in contrast to the others) while it offers significantly more detailed infrastructure level behaviour.

This experiment was also used to \emph{validate the results of DISSECT-CF} through a comparison with the other simulators. As the first step of validation, the simulator reported completion time of the last task in every trace specific measurement was collected (from $\mathcal{M}(k,\mathfrak{n},mc)$ -- where $\mathfrak{n}$ was set between one and a million, and $mc$ between one and five hundred). Then, these task completion times were compared with each other. The median of the difference between the simulators was less than 0.001\% (the average difference from the median was 2.15\%). The biggest differences occurred with simulated infrastructures containing the lowest number of physical machines: in such cases the median of difference between the simulators jumped to a little more than 0.29\% (with a sample standard deviation of 7.21\%). The reason behind such a great deviation is the magnified effect of virtual machine instantiation simulation (i.e., as VMs can only be instantiated on a single PM, they are mostly created and destroyed in a serial fashion).

\paragraph{The impact of energy metering}

So far, all presented measurements were executed without energy meters attached to the resource spreaders of DISSECT-CF. As energy metering is not available in GroudSim, the other two simulators would have suffered a disadvantage because of their ongoing metering simulation. As DISSECT-CF was consequently a better performer than the other two simulators, its further evaluation was focused on how one could set up energy metering in order to get comparable performance to the other simulators' meter-less operation.

\begin{figure}[tb]
\centering
\includegraphics[width=\columnwidth, trim=2.2cm 15.5cm 3.9cm 2.8cm, clip=true]{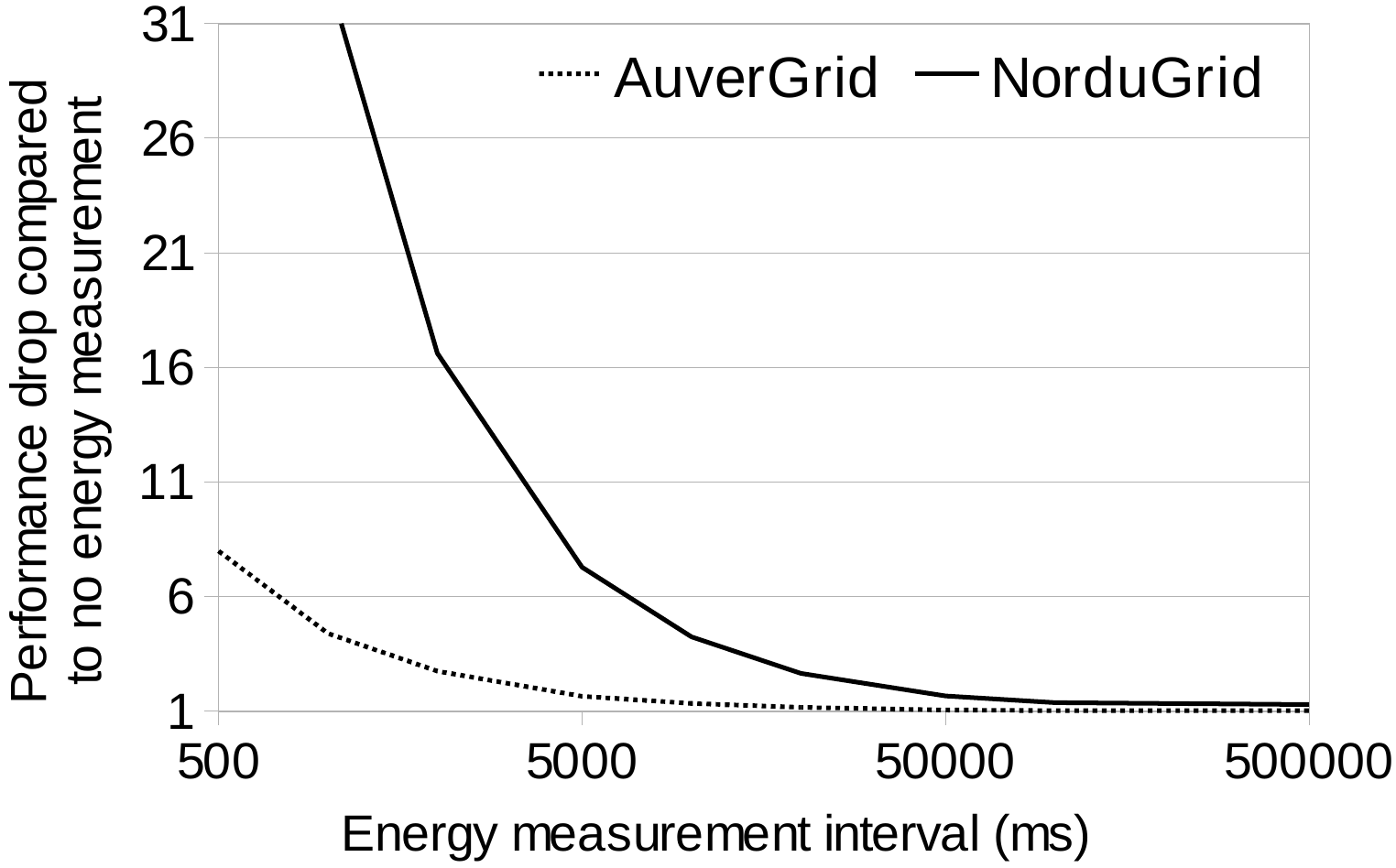}
\caption{Performance degradation in relation to energy metering frequency\label{FIG-EnergyPerformanceDrop}}
\end{figure}

Figure~\ref{FIG-EnergyPerformanceDrop} analyses the performance drop of DISSECT-CF observed when energy metering was requested for the complete simulated infrastructure with 20 physical machines. Just as before, the function $\mathcal{M}(DS, 1000000, 20)$ was evaluated, but now instead of comparing DISSECT-CF to other simulators, the comparison was made between the newly received values to values from the previously presented simulations without energy metering. Adding energy metering leads to significant performance differences between the various GWA traces. To show what is the possible performance degradation range in the various traces, the two extreme cases (the best and worst performing ones) are revealed in the figure. The presented cases show that selecting the energy measurement interval is crucial for well performing simulations. In general it is not recommended to use energy measurement intervals below one minute for long running simulations -- i.e., where the trace's length is over a few thousand tasks. If higher precision is needed, then selective metering techniques should be used that evaluate energy meters for only the necessary parts of the infrastructure; otherwise, one could experience a slowdown of over 50 times compared to meter-less runtimes.

\begin{figure}[tb]
\includegraphics[width=\columnwidth, trim=2.3cm 15.7cm 4.2cm 3.1cm, clip=true]{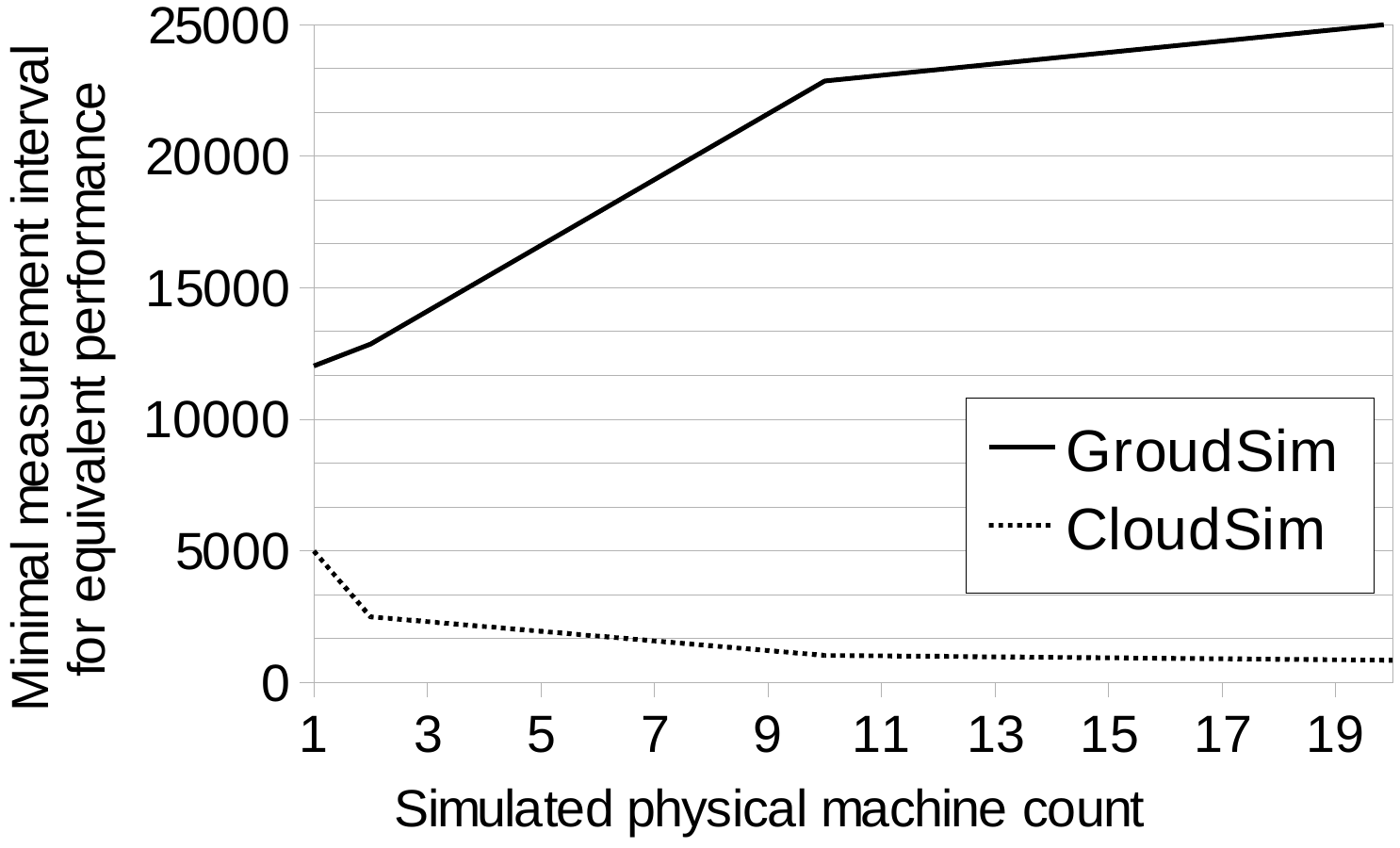}
\caption{Performance equivalence of energy metered DISSECT-CF simulations\label{FIG-EnergyEquivalence}}
\end{figure}
Finally, in Figure~\ref{FIG-EnergyEquivalence}, a comparison is shown about the performance drop caused by the metering to the meter-less setups of other simulators. The figure shows what the metering interval is -- despite the resulting performance drop -- that still provides equivalent performance to other simulators. As the figure shows, the performance drop also depends on the number of physical machines in the simulated infrastructure. Overall, if energy metering is desired on the entire simulated infrastructure, the metering interval must be set to at least five-seconds to receive comparable performance to CloudSim. In contrast, GroudSim's performance can be matched with a metering interval around 25-30 seconds. Anything above these values will give a performance advantage to DISSECT-CF. On the other hand, metering intervals below these values could still cause a better performing DISSECT-CF but this is highly dependent on the particular trace or virtual machine request pattern used during simulations.

\section{Conclusion} \label{sec-conclusion}
This article outlined several IaaS related schedulers that could be further improved with the use of cloud simulators. Then, it has shown that current simulators barely meet the demands of the scheduling oriented researchers: $(i)$ they often limit the accessibility of information, $(ii)$ they often hide the internal details of IaaS systems, $(iii)$ they frequently perform poorly in large-scale simulations, and $(iv)$ they provide scarcely any options for introducing new energy management techniques inside IaaS clouds. To overcome these issues, the article has proposed a new simulator called DISSECT-CF.

The proposed simulator targets information accessibility issues with open APIs and monitoring and performance related customisable events. The internal details of the cloud systems are also accessible, allowing simulation developers: $(i)$ to construct clouds in novel ways (e.g., introduction of new physical machine - virtual machine interaction techniques or cloud organisation topologies), and $(ii)$ to experiment with new cloud side behaviour (e.g., new VM schedulers, power states). DISSECT-CF utilises a new unified resource sharing mechanism that allows centralised performance optimisations and ensures scaling independently of the size of the infrastructure and the amount of tasks processed by the simulated system. Finally, the new simulator deeply integrates energy metering techniques (ranging from resource usage counters and energy consumption models to meter aggregators). These techniques not only allow further extensions but they allow selective and composite power metering to ensure minimal performance drops during the metering sessions.

The new simulator was evaluated by comparing it to small-scale but real-life environments. During this evaluation, it was presented how one should model CPU and memory intensive tasks, and it was also shown that the accuracy of the simulator's energy metering technique is also dependent on the new unified resource sharing mechanism. Experiments shown that the relative error of the new resource sharing technique is around 1\% in most cases. After the small-scale evaluation, DISSECT-CF was  compared with two state-of-the-art simulators (namely CloudSim and GroudSim). The comparisons were  focused on the scaling and performance characteristics of the simulators. The results revealed that these simulators often produce inaccurate results, while the new simulator not only provided more accurate outputs but also offered better performance.

Future research will consider several areas. First, there are plans to look how the simulator's resource sharing accuracy could be improved with stochastic sharing models and new low-level scheduling techniques. Next, support for more complex network topologies and additional node types (e.g., routers and switches) will also be prepared. As the current simulator merely provides raw data for user-side schedulers, an investigation of the necessary constructs and optimisations to better support the development and analysis of these schedulers also needs to be done. Finally, memory behaviour is practically neglected in the current simulator, because applications and openly available traces do not offer any details on this. In the future, a task level complex memory model is expected to be delivered that not only considers memory bandwidth utilisation but also access patterns. With this model the simulator's target could include accurate live migration support.

\section*{Software availability}
This article described the behaviour and features of DISSECT-CF version 0.9.5. The source code of the simulator is open and available (under the licensing terms of the GNU Lesser General Public License 3) at the following website: 

https://github.com/kecskemeti/dissect-cf. 

\section*{Acknowledgements}

This work was partially supported by European Union's Horizon 2020 research and innovation programme under grant agreement No 644179 (ENTICE) as well as by the COST Program Action IC1305: Network for Sustainable Ultrascale Computing (NESUS), finally the research has also received partial funding from the Austrian Science Fund project TRP 237-N23.


\begin{thebibliography}{10}

\bibitem{BERKELEYCLOUD}
M.~Armbrust, A.~Fox, R.~Griffith, et~al., Above the clouds: A berkeley view of
  cloud computing, Tech. Rep. UCB/EECS-2009-28, University of California at
  Berkley (February 2009).

\bibitem{BreakClouds}
L.~M. Vaquero, L.~Rodero-Merino, J.~Caceres, et~al., A break in the clouds:
  towards a cloud definition, SIGCOMM Computer Communication Review 39 (2008)
  50--55.
\newblock doi:10.1145/1496091.1496100.

\bibitem{SimuComp-AhmedSabayasachi}
A.~Ahmed and A.S. Sabyasachi.
\newblock Cloud computing simulators: A detailed survey and future direction.
\newblock In {\em Advance Computing Conference (IACC), 2014 IEEE
	International}, pages 866--872. IEEE, February 2014.
\newblock doi:10.1109/IAdCC.2014.6779436.

\bibitem{SimuCompare-zhao2012modeling}
W.~Zhao, Y.~Peng, F.~Xie, Z.~Dai, Modeling and simulation of cloud computing: A
  review, in: IEEE Asia Pacific Cloud Computing Congress (APCloudCC), IEEE,
  Shenzhen, 2012, pp. 20--24.
\newblock doi:10.1109/APCloudCC.2012.6486505.

\bibitem{SimGrid-Hirofuchi2013}
T.~Hirofuchi, A.~L{\`e}bre, Adding virtual machine abstractions into {SimGrid}:
  A first step toward the simulation of infrastructure-as-a-service concerns,
  in: Third International Conference on Cloud and Green Computing (CGC), IEEE,
  Karlsruhe, Germany, 2013, pp. 175--180.
\newblock doi:10.1109/CGC.2013.33.

\bibitem{iCanCloud-Nunez2011}
A.~N{\'u}{\~n}ez, J.~L. V{\'a}zquez-Poletti, A.~C. Caminero, J.~Carretero,
  I.~M. Llorente, Design of a new cloud computing simulation platform, in:
  Computational Science and Its Applications-ICCSA 2011, Vol. 6784 of Lecture
  Notes in Computer Science, Springer, Santander, Spain, 2011, pp. 582--593.
\newblock doi:10.1007/978-3-642-21931-3\_45.

\bibitem{DCWorms-kurowski2013dcworms}
K.~Kurowski, A.~Oleksiak, W.~Pi{k{a}}tek, T.~Piontek, A.~Przybyszewski,
  J.~W{k{e}}glarz, {DCworms}--a tool for simulation of energy efficiency in
  distributed computing infrastructures, Simulation Modelling Practice and
  Theory 39 (2013) 135--151.
\newblock doi:10.1016/j.simpat.2013.08.007.

\bibitem{GreenCloud-5683561}
D.~Kliazovich, P.~Bouvry, Y.~Audzevich, S.~Khan, {GreenCloud}: A packet-level
  simulator of energy-aware cloud computing data centers, in: Global
  Telecommunications Conference (GLOBECOM 2010), 2010, pp. 1--5.
\newblock doi:10.1109/GLOCOM.2010.5683561.

\bibitem{CloudSim-calheiros2011CloudSim}
R.~N. Calheiros, R.~Ranjan, A.~Beloglazov, C.~A. De~Rose, R.~Buyya, {CloudSim}:
  a toolkit for modeling and simulation of cloud computing environments and
  evaluation of resource provisioning algorithms, Software: Practice and
  Experience 41~(1) (2011) 23--50.
\newblock doi:10.1002/spe.995.

\bibitem{GroudSim-ostermann2011groudsim}
S.~Ostermann, K.~Plankensteiner, R.~Prodan, T.~Fahringer, Groudsim: an
  event-based simulation framework for computational grids and clouds, in:
  Euro-Par 2010 Parallel Processing Workshops, Vol. 6586 of Lecture Notes in
  Computer Science, Springer, 2011, pp. 305--313.
\newblock doi:10.1007/978-3-642-21878-1\_38.

\bibitem{SimGrid-velho2013validity}
P.~Velho, L.~M. Schnorr, H.~Casanova, A.~Legrand, On the validity of flow-level
  tcp network models for grid and cloud simulations, ACM Transactions on
  Modeling and Computer Simulation (TOMACS) 23~(4) (2013) 23.
\newblock doi:10.1145/2517448.

\bibitem{SimGrid-velho2009accuracy}
P.~Velho, A.~Legrand, Accuracy study and improvement of network simulation in
  the {SimGrid} framework, in: Proceedings of the 2nd International Conference
  on Simulation Tools and Techniques, ICST (Institute for Computer Sciences,
  Social-Informatics and Telecommunications Engineering), 2009, p.~13.
\newblock doi:10.4108/ICST.SIMUTOOLS2009.5592.

\bibitem{Iosup08thegrid}
A.~Iosup, H.~Li, M.~Jan, S.~Anoep, C.~Dumitrescu, L.~Wolters, D.~H.~J. Epema,
  The grid workloads archive, Future Generation Computer Systems 24~(7) (2008)
  672--686.
\newblock doi:10.1016/j.future.2008.02.003.

\bibitem{Sobie:2013:HSC:2465848.2465850}
R.~Sobie, A.~Agarwal, I.~Gable, C.~Leavett-Brown, M.~Paterson, R.~Taylor,
  A.~Charbonneau, R.~Impey, W.~Podiama, {HTC} scientific computing in a
  distributed cloud environment, in: Proceedings of the 4th ACM Workshop on
  Scientific Cloud Computing, Science Cloud '13, ACM, New York, NY, USA, 2013,
  pp. 45--52.
\newblock doi:10.1145/2465848.2465850.

\bibitem{6332105}
H.~Song, J.~Li, X.~Liu, {IdleCached}: An idle resource cached dynamic
  scheduling algorithm in cloud computing, in: 9th International Conference on
  Ubiquitous Intelligence Computing and 9th International Conference on
  Autonomic Trusted Computing (UIC/ATC), 2012, pp. 912--917.
\newblock doi:10.1109/UIC-ATC.2012.24.

\bibitem{rodrigez2009}
M.~Rodr{\'\i}guez, D.~Tapiador, J.~Font{\'a}n, E.~Huedo, R.~Montero,
  I.~Llorente, Dynamic provisioning of virtual clusters for grid computing, in:
  E.~C{\'e}sar, M.~Alexander, A.~Streit, J.~Tr{\"a}ff, C.~C{\'e}rin,
  A.~Kn{\"u}pfer, D.~Kranzlm{\"u}ller, S.~Jha (Eds.), Euro-Par 2008 Workshops -
  Parallel Processing, Vol. 5415 of Lecture Notes in Computer Science, Springer
  Berlin Heidelberg, 2009, pp. 23--32.
\newblock doi:10.1007/978-3-642-00955-6\_4.

\bibitem{Sotomayor:2008:CBE:1383422.1383434}
B.~Sotomayor, K.~Keahey, I.~Foster, Combining batch execution and leasing using
  virtual machines, in: Proceedings of the 17th International Symposium on High
  Performance Distributed Computing, HPDC '08, ACM, New York, NY, USA, 2008,
  pp. 87--96.
\newblock doi:10.1145/1383422.1383434.

\bibitem{Tordsson2012358}
J.~Tordsson, R.~S. Montero, R.~Moreno-Vozmediano, I.~M. Llorente, Cloud
  brokering mechanisms for optimized placement of virtual machines across
  multiple providers, Future Generation Computer Systems 28~(2) (2012) 358 --
  367.
\newblock doi:10.1016/j.future.2011.07.003.

\bibitem{5279590}
E.~Elmroth, L.~Larsson, Interfaces for placement, migration, and monitoring of
  virtual machines in federated clouds, in: Eighth International Conference on
  Grid and Cooperative Computing. GCC '09., 2009, pp. 253--260.
\newblock doi:10.1109/GCC.2009.36.

\bibitem{tsakalozos2011}
K.~Tsakalozos, M.~Roussopoulos, A.~Delis, {VM} placement in non-homogeneous
  {IaaS}-clouds, in: G.~Kappel, Z.~Maamar, H.~Motahari-Nezhad (Eds.),
  Service-Oriented Computing, Vol. 7084 of Lecture Notes in Computer Science,
  Springer Berlin Heidelberg, 2011, pp. 172--187.
\newblock doi:10.1007/978-3-642-25535-9\_12.

\bibitem{6009246}
D.~Jayasinghe, C.~Pu, T.~Eilam, M.~Steinder, I.~Whally, E.~Snible, Improving
  performance and availability of services hosted on {IaaS} clouds with
  structural constraint-aware virtual machine placement, in: IEEE International
  Conference on Services Computing (SCC), 2011, pp. 72--79.
\newblock doi:10.1109/SCC.2011.28.

\bibitem{5958802}
R.~Ghosh, V.~Naik, K.~Trivedi, Power-performance trade-offs in {IaaS} cloud: A
  scalable analytic approach, in: IEEE/IFIP 41st International Conference on
  Dependable Systems and Networks Workshops (DSN-W), 2011, pp. 152--157.
\newblock doi:10.1109/DSNW.2011.5958802.

\bibitem{6253507}
E.~Feller, C.~Rohr, D.~Margery, C.~Morin, Energy management in iaas clouds: A
  holistic approach, in: IEEE 5th International Conference on Cloud Computing
  (CLOUD), 2012, pp. 204--212.
\newblock doi:10.1109/CLOUD.2012.50.

\bibitem{Wang20131661}
Lizhe Wang, Samee~U. Khan, Dan Chen, Joanna Kolodziej, Rajiv Ranjan, Cheng
zhong Xu, and Albert Zomaya.
\newblock Energy-aware parallel task scheduling in a cluster.
\newblock {\em Future Generation Computer Systems}, 29(7):1661 -- 1670,
September 2013.
\newblock Including Special sections: Cyber-enabled Distributed Computing for
Ubiquitous Cloud and Network Services \&amp; Cloud Computing and Scientific
Applications --- Big Data, Scalable Analytics, and Beyond.
\newblock doi:10.1016/j.future.2013.02.010.

\bibitem{Ongaro:2008:SIV:1346256.1346258}
D.~Ongaro, A.~L. Cox, S.~Rixner, Scheduling {I/O} in virtual machine monitors,
  in: Proceedings of the Fourth ACM SIGPLAN/SIGOPS International Conference on
  Virtual Execution Environments, VEE '08, ACM, New York, NY, USA, 2008, pp.
  1--10.
\newblock doi:10.1145/1346256.1346258.

\bibitem{5289182}
G.~von Laszewski, L.~Wang, A.~Younge, X.~He, Power-aware scheduling of virtual
  machines in {DVFS}-enabled clusters, in: IEEE International Conference on
  Cluster Computing and Workshops. CLUSTER '09., 2009, pp. 1--10.
\newblock doi:10.1109/CLUSTR.2009.5289182.

\bibitem{Wong:2008:TAF:1400097.1400102}
C.~S. Wong, I.~Tan, R.~D. Kumari, F.~Wey, Towards achieving fairness in the
  linux scheduler, SIGOPS Oper. Syst. Rev. 42~(5) (2008) 34--43.
\newblock doi:10.1145/1400097.1400102.

\bibitem{4510751}
J.~Yang, X.~Zhou, M.~Chrobak, Y.~Zhang, L.~Jin, Dynamic thermal management
  through task scheduling, in: IEEE International Symposium on Performance
  Analysis of Systems and software. ISPASS 2008., 2008, pp. 191--201.
\newblock doi:10.1109/ISPASS.2008.4510751.

\bibitem{CloudSim-buyya2009modeling}
R.~Buyya, R.~Ranjan, R.~N. Calheiros, Modeling and simulation of scalable cloud
  computing environments and the {CloudSim} toolkit: Challenges and
  opportunities, in: International Conference on High Performance Computing \&
  Simulation (HPCS'09)., IEEE, Leipzig, 2009, pp. 1--11.
\newblock doi:10.1109/HPCSIM.2009.5192685.

\bibitem{GridSIM-buyya2002gridsim}
R.~Buyya, M.~Murshed, Gridsim: A toolkit for the modeling and simulation of
  distributed resource management and scheduling for grid computing,
  Concurrency and Computation: Practice and Experience 14~(13-15) (2002)
  1175--1220.
\newblock doi:10.1002/cpe.710.

\bibitem{CloudSim-garg2011networkCloudSim}
S.~K. Garg, R.~Buyya, {NetworkCloudSim}: modelling parallel applications in
  cloud simulations, in: Fourth IEEE International Conference on Utility and
  Cloud Computing (UCC), IEEE, Victoria, NSW, 2011, pp. 105--113.
\newblock doi:10.1109/UCC.2011.24.

\bibitem{CloudSim-beloglazov2012optimal}
A.~Beloglazov, R.~Buyya, Optimal online deterministic algorithms and adaptive
  heuristics for energy and performance efficient dynamic consolidation of
  virtual machines in cloud data centers, Concurrency and Computation: Practice
  and Experience 24~(13) (2012) 1397--1420.
\newblock doi:10.1002/cpe.1867.

\bibitem{CloudSimExt-DVFS}
Tom Gu\'erout, Thierry Monteil, Georges~Da Costa, Rodrigo~Neves Calheiros,
Rajkumar Buyya, and Mihai Alexandru.
\newblock Energy-aware simulation with \{DVFS\}.
\newblock {\em Simulation Modelling Practice and Theory}, 39:76 -- 91, December
2013.
\newblock S.I. Energy efficiency in grids and clouds.
\newblock doi:10.1016/j.simpat.2013.04.007.

\bibitem{CloudSimExt-li2012dartcsim}
X.~Li, X.~Jiang, P.~Huang, K.~Ye, {DartCSim}: An enhanced user-friendly cloud
  simulation system based on {CloudSim} with better performance, in: 2nd
  International Conference on Cloud Computing and Intelligent Systems (CCIS),
  Vol.~1, IEEE, Hangzhou, 2012, pp. 392--396.
\newblock doi:10.1109/CCIS.2012.6664434.

\bibitem{SIMIC-6531742}
S.~Sotiriadis, N.~Bessis, N.~Antonopoulos, A.~Anjum, {SimIC}: Designing a new
  inter-cloud simulation platform for integrating large-scale resource
  management, in: 27th International Conference on Advanced Information
  Networking and Applications (AINA), 2013, pp. 90--97.
\newblock doi:10.1109/AINA.2013.123.

\bibitem{SIMIC-6550488}
S.~Sotiriadis, N.~Bessis, N.~Antonopoulos, Towards inter-cloud simulation
  performance analysis: Exploring service-oriented benchmarks of clouds in
  {SimIC}, in: 27th International Conference on Advanced Information Networking
  and Applications Workshops (WAINA), 2013, pp. 765--771.
\newblock doi:10.1109/WAINA.2013.196.

\bibitem{CloudSimExt-shi2011energy}
Y.~Shi, X.~Jiang, K.~Ye, An energy-efficient scheme for cloud resource
  provisioning based on {CloudSim}, in: IEEE International Conference on
  Cluster Computing (CLUSTER), IEEE, Austin, TX, 2011, pp. 595--599.
\newblock doi:10.1109/CLUSTER.2011.63.

\bibitem{CloudSimExt-SLA}
Alexandru-Florian Antonescu and Torsten Braun.
\newblock Sla-driven simulation of multi-tenant scalable cloud-distributed
enterprise information systems.
\newblock In Florin Pop and Maria Potop-Butucaru, editors, {\em Adaptive
	Resource Management and Scheduling for Cloud Computing}, Lecture Notes in
Computer Science, pages 91--102. Springer International Publishing, November
2014.
\newblock doi:10.1007/978-3-319-13464-2\_7.

\bibitem{CloudSimExt-jararweh2012teachcloud}
Y.~Jararweh, Z.~Alshara, M.~Jarrah, M.~Kharbutli, M.~Alsaleh, Teachcloud: a
  cloud computing educational toolkit, Int. J. of Cloud Computing 2~(2/3)
  (2012) 237--257.
\newblock doi:10.1504/IJCC.2013.055269.

\bibitem{CloudSimExt-wickremasinghe2010cloudanalyst}
B.~Wickremasinghe, R.~N. Calheiros, R.~Buyya, Cloudanalyst: A {CloudSim}-based
  visual modeller for analysing cloud computing environments and applications,
  in: 24th IEEE International Conference on Advanced Information Networking and
  Applications (AINA), IEEE, 2010, pp. 446--452.
\newblock doi:10.1109/AINA.2010.32.

\bibitem{CloudSimExt-CloudReports}
Thiago Teixeira~S\'a, Rodrigo~N. Calheiros, and Danielo~G. Gomes.
\newblock {CloudReports: An Extensible Simulation Tool for Energy-Aware Cloud
	Computing Environments}.
\newblock In Zaigham Mahmood, editor, {\em Cloud Computing}, Computer
Communications and Networks, pages 127--142. Springer International
Publishing, October 2014.
\newblock doi:10.1007/978-3-319-10530-7\_6.

\bibitem{SimGrid-casanova2001simgrid}
H.~Casanova, {SimGrid}: A toolkit for the simulation of application scheduling,
  in: First IEEE/ACM International Symposium on Cluster Computing and the Grid,
  IEEE, Brisbane, Qld., 2001, pp. 430--437.
\newblock doi:10.1109/CCGRID.2001.923223.

\bibitem{SimGrid-hirofuchi2013adding}
T.~Hirofuchi, A.~L{\`e}bre, L.~Pouilloux, et~al., Adding a live migration model
  into {SimGrid}, one more step toward the simulation of
  infrastructure-as-a-service concerns, in: 5th IEEE International Conference
  on Cloud Computing Technology and Science (IEEE CloudCom), Bristol, UK, 2013,
  pp. 96--103.
\newblock doi:10.1109/CloudCom.2013.20.

\bibitem{GreenCloud-kliazovich2012greencloud}
D.~Kliazovich, P.~Bouvry, S.~U. Khan, {GreenCloud}: a packet-level simulator of
  energy-aware cloud computing data centers, The Journal of Supercomputing
  62~(3) (2012) 1263--1283.
\newblock doi:10.1007/s11227-010-0504-1.

\bibitem{iCanCloud-nunez2011design}
A.~N{\'u}{\~n}ez, G.~Castane, J.~Vazquez-Poletti, A.~Caminero, J.~Carretero,
  I.~Llorente, Design of a flexible and scalable hypervisor module for
  simulating cloud computing environments, in: International Symposium on
  Performance Evaluation of Computer \& Telecommunication Systems (SPECTS),
  IEEE, 2011, pp. 265--270.

\bibitem{iCanCloud-nunez2012icancloud}
A.~N{\'u}{\~n}ez, J.~L. V{\'a}zquez-Poletti, A.~C. Caminero, G.~G.
  Casta{\~n}{\'e}, J.~Carretero, I.~M. Llorente, {iCanCloud}: A flexible and
  scalable cloud infrastructure simulator, Journal of Grid Computing 10~(1)
  (2012) 185--209.
\newblock doi:10.1007/s10723-012-9208-5.

\bibitem{GroudSim-ostermann2011integration}
S.~Ostermann, K.~Plankensteiner, D.~Bodner, G.~Kraler, R.~Prodan, Integration
  of an event-based simulation framework into a scientific workflow execution
  environment for grids and clouds, in: Towards a Service-Based Internet, Vol.
  6994 of Lecture Notes in Computer Science, Springer, Poznan, Poland, 2011,
  pp. 1--13.
\newblock doi:10.1007/978-3-642-24755-2\_1.

\bibitem{SPECI-sriram2009speci}
I.~Sriram, {SPECI}, a simulation tool exploring cloud-scale data centres, in:
  Cloud Computing, Vol. 5931 of Lecture Notes in Computer Science, Springer,
  2009, pp. 381--392.
\newblock doi:10.1007/978-3-642-10665-1\_35.

\bibitem{DCSim-tighe2012dcsim}
M.~Tighe, G.~Keller, M.~Bauer, H.~Lutfiyya, {DCSim}: A data centre simulation
  tool for evaluating dynamic virtualized resource management, in: 8th
  International Conference on Network and Service Management (CNSM), IEEE,
  2012, pp. 385--392.

\bibitem{DCWorms-piatek2013dcworms}
W.~Piatek, {DCworms}--a tool for simulation of energy efficiency in data
  centers, in: Energy Efficiency in Large Scale Distributed Systems, Vol. 8046
  of Lecture Notes in Computer Science, Springer, 2013, pp. 118--124.
\newblock doi:10.1007/978-3-642-40517-4\_11.

\bibitem{DCSim-tighe2013towards}
M.~Tighe, G.~Keller, J.~Shamy, M.~Bauer, H.~Lutfiyya, Towards an improved data
  centre simulation with {DCSim}, in: Proceedings of the 9th International
  Conference on Network and Service Management, CNSM 2013, IEEE, Zurich,
  Switzerland, 2013, pp. 364--372.
\newblock doi:10.1109/CNSM.2013.6727859.

\bibitem{MaxMinFair}
D.~Bertsekas, R.~Gallager (Eds.), Data Networks, {Second} Edition, Vol.~2,
  Prentice Hall International, Englewood Cliffs, New Jersey 07632, 1992, Ch. 6.
  Flow Control, pp. 493--536.

\bibitem{opennebula2011}
D.~Miloji{\v{c}}i{\'c}, I.~M. Llorente, R.~S. Montero, {OpenNebula}: {A} cloud
  management tool, Internet Computing, IEEE 15~(2) (2011) 11--14.
\newblock doi:10.1109/MIC.2011.44.

\bibitem{DISS}
G.~Kecskemeti, Foundations of Efficient Virtual Appliance Based Service
  Deployments: New Techniques for Virtual Appliance Delivery and Size
  Optimization in Infrastructure as a Service Clouds, ISBN: 978-3-8484-0383-7,
  LAP Lambert Academic Publishing, Saarbr{\"u}cken, 2012.

\end{thebibliography}
\end{document}